\documentclass[12pt,a4paper]{article}
\usepackage{jheppub}
\pdfoutput = 1

\usepackage{amsmath}
\usepackage{amssymb}
\usepackage{bbm}
\usepackage{bbold}
\usepackage{braket}
\usepackage{caption}
\usepackage{color}
    \definecolor{darkgreen}{rgb}{0,0.5,0}
    \definecolor{darkred}{rgb}{0.5,0,0}
    \definecolor{darkblue}{rgb}{0,0,0.6}
    \definecolor{purple}{rgb}{0.4,.2,0.7}

\usepackage{float}
\usepackage{graphicx}
\usepackage{hyperref}

\usepackage{mathabx}
\usepackage{mathtools}
\usepackage{pdfsync}
\usepackage{slashed}
\usepackage[normalem]{ulem}
\usepackage{upgreek}
\usepackage{url}

\newcommand\Dbar{\check{D}}

\usepackage[ragged]{footmisc}
\def\be{\begin{equation}}
\def\ee{\end{equation}}

\renewcommand{\tilde}{\widetilde}

\makeatletter
\newcommand{\vast}{\bBigg@{4}}
\newcommand{\Vast}{\bBigg@{5}}
\makeatother

\numberwithin{equation}{section}

\begin{document}
\title{A deformed IR: a new IR fixed point for four-dimensional holographic  theories}

\author[a]{Gary~T.~Horowitz,}
\author[b]{Maciej Kolanowski,}
\author[c]{Jorge~E.~Santos}
\affiliation[a]{Department of Physics, University of California at Santa Barbara, Santa Barbara, CA 93106, U.S.A.}
\affiliation[b]{Institute of Theoretical Physics, Faculty of Physics, University of Warsaw, Pasteura 5, 02-093 Warsaw, Poland}
\affiliation[c]{Department of Applied Mathematics and Theoretical Physics, University of Cambridge, Wilberforce Road, Cambridge, CB3 0WA, UK}

\emailAdd{horowitz@ucsb.edu}
\emailAdd{maciej.kolanowski@fuw.edu.pl}
\emailAdd{jss55@cam.ac.uk}

\abstract{
In holography, the IR behavior of a quantum system at nonzero  density is described by the near horizon geometry of an extremal charged black hole. It is commonly believed that for systems on $S^3$, this near horizon geometry is $AdS_2\times S^3 $. We show that this is not the case: generic static, nonspherical perturbations of $AdS_2\times S^3 $ blow up at the horizon, showing that it is not a stable IR fixed point. We then construct a new near horizon geometry which is invariant under only $SO(3)$ (and not $SO(4)$) symmetry and show that it is stable to $SO(3)$-preserving perturbations (but not in general). We also show that an open set of nonextremal, $SO(3)$-invariant charged black holes develop this new near horizon geometry in the limit $T \to 0$. Our new IR geometry still has $AdS_2$ symmetry, but it is warped over a deformed sphere.  We also construct many other near horizon geometries, including some with no rotational symmetries, but expect them all to be unstable IR fixed points. 
}

\maketitle
\section{Introduction} 

A standard entry in the holographic dictionary states that the dual of a thermal state of a field theory at temperature $T$ and chemical potential $\mu$ is  described by an asymptotically anti-de Sitter (AdS) charged black hole  \cite{Hartnoll:2011fn}. If the field theory is on a round sphere and $\mu$ is constant, the
black hole is given by the Reissner-Nordstr\"om (RN) AdS  solution. Since another tenet of holography is that the radial direction corresponds to an energy scale in the field theory \cite{Peet:1998wn}, the IR behavior of the theory is described by the near horizon limit of the extremal\footnote{By extremal, we will always refer to the $T=0$ solution. See \cite{Dias:2021vve} for an example where this is not the case.}  solution, which for RN AdS is $AdS_2\times S^n$.

In four bulk dimensions, $AdS_2\times S^2$  remains the near horizon geometry of the extremal black hole even if one deforms the chemical potential or the boundary metric to a static, nonspherical configuration \cite{Horowitz:2022mly}. (For smooth horizons, it has been shown that the  only static near horizon solutions in Einstein-Maxwell theory are $AdS_2\times H$ where $H$ is a space of constant curvature, i.e., a sphere, torus, or compact Riemann surface \cite{Kunduri:2013gce}. Even though this theorem does not apply to generic nonspherical solutions since the horizon is singular \cite{Horowitz:2022mly}, the conclusion still holds.) Intuitively, this is because the extremal horizon is infinitely far away from any effect outside the horizon (along a static surface). From the dual field theory perspective, $AdS_2\times S^2$ describes a stable IR fixed point. Note that we are not referring to dynamical stability, but rather stability in the RG sense, which is a property of the space of static solutions.  It is widely believed that in higher dimensions, $AdS_2\times S^n$ similarly describes a stable IR fixed point.

\begin{figure}[th]
\centering \includegraphics[width=0.8\textwidth]{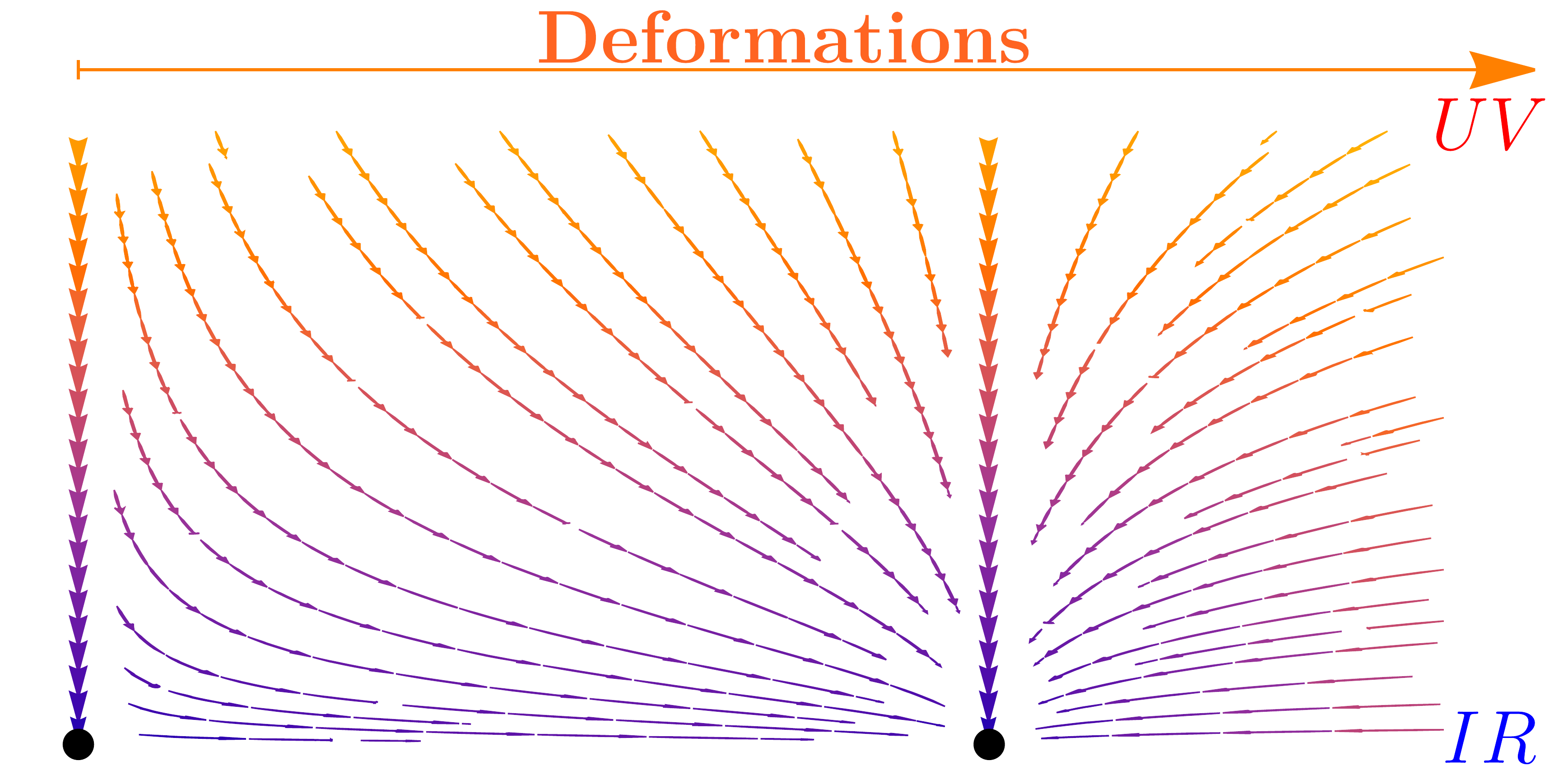}
\caption{\label{fig:rg_flow} Illustration of stable and unstable fixed points in the RG sense.
}
\end{figure}

We will show that this common belief is incorrect. For all $n> 2$, generic static nonspherical perturbations of $AdS_2\times S^n$ blow up on the horizon, even though the horizon is still infinitely far away. We will construct a new near horizon geometry in $D= n+2=5$ that is invariant under only $SO(3)$ (and not $SO(4)$) symmetry and show that it is stable to $SO(3)$-preserving perturbations. In addition, we construct an open set of nonextremal, $SO(3)$-invariant  black holes and show that as $T\to 0$, they approach our new near horizon geometry. This shows that, within this symmetry class, our new solution is a stable IR fixed point for four-dimensional holographic theories. Of course $SO(4)$-symmetry is a special point in our class, and if one imposes it, one still flows to $AdS_2\times S^3$, but this is now seen as an unstable fixed point. This is illustrated in Fig.~\ref{fig:rg_flow}. 

There is actually a one-parameter family of these new IR geometries which are conveniently labelled by the total charge $Q$. While we do not have analytic expressions for the new solutions, we can construct them numerically and (for small $Q$) check them with an analytic perturbative expansion.  When $Q$ is small, the solutions are close to $AdS_2 \times S^3$. However, as $Q$ increases, the curvature near the poles of the $S^3$ decreases so the sphere becomes flattened. For even larger $Q$, this curvature  becomes negative. In the limit $Q\to \infty$ the curvature near the poles approaches a finite negative value, so the sphere looks like two large hyperbolic disks joined by a positive curvature ring around the equator. There is still an $AdS_2$ factor, but now it is warped as one moves around the deformed sphere.

The solutions we find turn out to be unstable to perturbations that break the $SO(3)$ symmetry, so they do not describe true stable IR fixed points.
It is an important open problem to find the gravitational description of these true stable fixed points.  One might think that a reasonable approach to this problem is to first  classify all possible near horizon geometries in higher dimensions, and then study their stability. However  this approach is doomed to failure since we expect there to be an infinite number of near horizon geometries. We show how to construct large families of them, including some with no rotational symmetries at all. Unfortunately, the solutions we construct are all RG unstable.

\section{\label{sec:unstable}Reissner-Nordstr\"om AdS is IR unstable in $D=5$}

To study IR fixed points of four-dimensional holographic theories, we work with the $D=5$ Einstein-Maxwell theory
\begin{equation}
S = \frac{1}{16\pi G_5}\int_{\mathcal{M}} \mathrm{d}^5x\sqrt{-g}\left(R-F_{ab}F^{ab}+\frac{12}{L^2}\right)+S_{\partial \mathcal{M}}\,,
\end{equation}
where $F=\mathrm{d}A$, $A$ is the Maxwell 1-form potential, $L$ is the $AdS_5$ length scale and $G_5$ is the five-dimensional Newton's constant. The equations of motion derived from this action read
\begin{subequations}
\begin{equation}
R_{ab}-\frac{R}{2}g_{ab}-\frac{6}{L^2} g_{ab}=2\left(F_{a}^{\phantom{a}c}F_{bc}-\frac{g_{ab}}{4}F_{cd}F^{cd}\right)
\end{equation}
and
\begin{equation}
\nabla^a F_{ab}=0\,.
\end{equation}
\label{eq:einsteinmaxwell}
\end{subequations}

There is a unique two-parameter spherical solution to these equations (with a non-constant areal radius) which is given by
\begin{subequations}
\begin{align}
&\mathrm{d}s^2_{\mathrm RN} = -f(r)\mathrm{d}t^2+\frac{\mathrm{d}r^2}{f(r)}+r^2 \mathrm{d}\Omega_3^2\,,
\\
& A_{\mathrm RN}=\frac{q}{r_+^2}\left(1-\frac{r_+^2}{r^2}\right)\,\mathrm{d}t\,,
\end{align}
\end{subequations}
where $\mathrm{d}\Omega_3^2$ is the metric on a unit radius round three-sphere and
\begin{equation}
f(r)=\frac{r^2}{L^2}+1-\frac{r_+^2}{r^2}\left(\frac{r_+^2}{L^2}+1+\frac{4 q^2}{3 r_+^4}\right)+\frac{4 q^2}{3 r^4}\,.
\end{equation}
This is the familiar Reissner-Nordstr\"om AdS solution (RN AdS).

We will take $|q|\leq q_{\rm ext}$, with
\begin{equation}
q_{\rm ext}=\frac{\sqrt{3}}{2}\sqrt{1+2\frac{r_+^2}{L^2}}r_+^2 \,,
\end{equation}
so that $r=r_+$ is the largest root of $f(r)=0$. The black hole event horizon is then the null hypersurface $r=r_+$, where $f(r)$ vanishes. For $|q|<q_{\rm ext}$, $f(r)$ vanishes linearly and the black hole has non-vanishing Hawking temperature
\begin{equation}
T_H=\frac{|f^\prime(r_+)|}{4\pi}=\frac{2\,q_{\rm ext}^2}{3\,\pi\,r_+^4}\left(1-\frac{q^2}{q_{\rm ext}^2}\right)\,,
\end{equation}
whereas for $q=q_{\rm ext}$, $f(r)$ vanishes quadratically at $r=r_+$ and the hole is said to be extremal and has vanishing temperature. The parameter $q$ determines the total charge of the black hole by
\begin{equation}
Q=\frac{\pi\,q}{G}\,,
\end{equation}
while its energy $E$, chemical potential $\mu$ and entropy $S$ are given by
\begin{equation}
E=\frac{3 \pi r_+^2}{8 G}\left(\frac{r_+^2}{L^2}+1+\frac{4 q^2}{3 r_+^4}\right)\,,\quad \mu = \frac{q}{r_+^2}\,,\quad\text{and}\quad S=\frac{\pi^2}{2G}r_+^3\,,
\end{equation}
respectively. It is a simple exercise to show that all of these thermodynamic quantities satisfy the first law of black hole mechanics
\begin{equation}
\mathrm{d}E=T_H\,\mathrm{d}S+\mu\,\mathrm{d}Q\,.
\end{equation}

Hereafter we will focus on the extremal case. In particular, we are interested in  the near horizon geometry of the RN AdS black hole. To obtain this, we take a limit where we zoom near the extremal horizon located at $r=r_+$. Define new coordinates $\rho, T$ by
\begin{equation}
r=r_+(1+\lambda \rho)\quad\text{and}\quad t = \frac{L_{\mathrm{AdS}_2}^2}{r_+}\frac{T}{\lambda}\,,
\end{equation}
where $\lambda$ is a constant and we defined
\begin{equation}
L_{\mathrm{AdS}_2}^2\equiv \frac{r_+^2 L^2 }{4(L^2+3 r_+^2)}\,.
\label{eq:AdS2radius}
\end{equation}
We then take the limit  $\lambda\to 0$. The resulting line element is the Robinson-Bertotti solution (or a five-dimensional version thereoff), which takes the familiar $AdS_2\times S^3$ form
\begin{subequations}
\label{eq:RB}
\begin{equation}
\mathrm{d}s^2_{\rm RB}=L_{\mathrm{AdS}_2}^2\left(-\rho^2\,\mathrm{d}T^2+\frac{\mathrm{d}\rho^2}{\rho^2}\right)+r_+^2\mathrm{d}\Omega_3^2\,,
\label{eq:robinsonbertotti}
\end{equation}
and
\begin{equation}
A_{\rm RB}=\frac{2q_{\rm ext}}{r_+^3}\,\rho\,L_{\mathrm{AdS}_2}^2\, \mathrm{d}T\,,
\end{equation}
\end{subequations}
with $\rho=0$ being the black hole horizon, which in this limit yields the $AdS_2$ Poincar\'e horizon. The Robinson-Bertotti solution is itself a solution of the Einstein-Maxwell equations, since it is just a particular limit of the RN AdS black hole.

This near horizon geometry, according to the standard rules of AdS/CFT \cite{DHoker:2002nbb}, controls the IR of the dual theory. For this reason zero temperature solutions such as the one above, are often called IR geometries.
To understand whether a given IR geometry is stable, in the RG sense, we perturb the IR geometry by time-independent perturbations $(h,a)$, where $h$ and $a$ are metric and gauge field perturbations, respectively.

One might think that perturbing (\ref{eq:RB}) is a complicated task, but it turns out that symmetry can help us. We first note that $AdS_2$ has constant curvature. This means we can use harmonic functions on $AdS_2$ as building blocks for constructing our generic perturbations $(h,a)$. For time independent perturbations, harmonic functions on $AdS_2$ take a particularly simple form:
\begin{equation}
\Box_{\mathrm{AdS}_2} \mathbb{S}_{\gamma}(\rho)-\frac{\gamma(\gamma+1)}{L_{\mathrm{AdS}_2}^2}\mathbb{S}_{\gamma}(\rho)=0\quad \Rightarrow\quad \mathbb{S}_{\gamma}(\rho)=C\,\rho^{\gamma}\,.
\end{equation}
where $C$ is a normalisation constant.

To construct perturbations $(h,a)$ we use $\mathbb{S}_{\gamma}(\rho)$ as building blocks.   Let $I$ be an $AdS_2$ index and $\hat{I}$ an index on $S^3$. It then follows that metric perturbations with indices on the $S^3$ only behave as scalars under coordinate transformations on $AdS_2$, so we take
\begin{equation}
h_{\hat{I}\hat{J}}=\mathbb{S}_{\gamma}(\rho)\,\hat{h}_{\hat{I}\hat{J}}
\end{equation}
where $\hat{h}_{\hat{I}\hat{J}}$ is a symmetric 2-tensor on $S^3$. The metric components $h_{I\hat{J}}$, on the other hand, behave as vectors, so we set
\begin{equation}
h_{I\hat{J}}={D}_{I}\mathbb{S}_{\gamma}(\rho)\,\hat{h}_{\hat{J}}\,
\end{equation}
where ${D}_{I}$ is the covariant derivative on $AdS_2$ and $\hat{h}_{\hat{J}}$ a vector on $S^3$.

Finally, we come to metric perturbations with indices on $AdS_2$. These behave as symmetric 2-tensors with respect to coordinate transformations on $AdS_2$. Any symmetric 2-tensor can be built from a trace and traceless symmetric 2-tensor. The latter two need to be built from $\mathbb{S}_{\gamma}(\rho)$. We thus set 
\begin{equation}
h_{IJ}=\mathbb{S}_\gamma\,\hat{h}_L\,{g}_{IJ}+\hat{h}_T \mathbb{S}_{IJ}
\end{equation}
with
\begin{equation}
\mathbb{S}_{IJ}\equiv {D}_{I}{D}_{J}\mathbb{S}_{\gamma}(\rho)-\frac{1}{2}\frac{\gamma(\gamma+1)}{L_{\mathrm{AdS}_2}^2}g_{IJ}\mathbb{S}_{\gamma}(\rho)
\end{equation}
where $g_{IJ}$ is the metric on $AdS_2$ and $\hat{h}_L$ and $\hat{h}_T$ are functions on $S^3$. 

We now discuss the thorny issue of gauge invariance. A generic gauge transformation $\xi$ can be written using the same procedure as above. In particular we find
\begin{equation}
\xi_a \mathrm{d}x^a = \hat{\xi}_S\,{D}_{I}\mathbb{S}_{\gamma}(\rho)\,\mathrm{d}x^{I}+\hat{\xi}_{\hat{I}}\,\mathbb{S}_{\gamma}(\rho)\,\mathrm{d}x^{\hat{I}}\,,
\end{equation}
where $\hat{\xi}_S$ is a scalar on $S^3$ and $\hat{\xi}_{\hat{I}}$ is a vector on $S^3$. The metric perturbations will transform under an infinitesimal transformation generated by $\xi$ according to
\begin{equation}
\delta h=\mathcal{L}_{\xi}  g_{\rm RB}\,,
\end{equation}
which induces 
\begin{subequations}
\begin{align}
\delta \hat{h}_L&=\frac{\gamma(\gamma+1)}{L_{\rm AdS_2}^2}\hat{\xi}_S
\\
\delta \hat{h}_T&=2\,\hat{\xi}_S
\\
\delta \hat h_{\hat{I}}&=2(\hat{D}_{\hat{I}}\hat{\xi}_S+\hat{\xi}_{\hat{I}})
\\
\delta h_{\hat{I}\hat{J}}&=(
\mathcal{L}_{\hat{\xi}}\hat{g})_{\hat{I}\hat{J}}
\end{align}
\end{subequations}%
where $\hat{g}_{\hat{I}\hat{J}}$ is the metric on a unit radius round $S^3$ and $\hat{D}_{\hat{I}}$ its metric preserving covariant derivative. We will work in a gauge where we choose $\hat{\xi}_S$ and $\hat{\xi}_{\hat{I}}$ so that $\hat{h}_T= \hat h_{\hat{I}}=0$.

For the gauge field perturbation, we have
\begin{equation}
a = \hat{a}_S\,\mathbb{S}_{\gamma}(\rho)\,\mathrm{d}T\,,
\end{equation}
where $\hat{a}_S$ is a function on $S^3$. Our perturbed gauge and metric field configurations are thus, in our gauge, parametrised by $\hat{h}_L$, $\hat{h}_{\hat{I}\hat{J}}$ and $\hat{a}_S$, which depend on the $S^3$ angles only.

Next we repeat the procedure and decompose the remaining perturbations in terms of spherical harmonics on the $S^3$.  These, in turn, are parametrised by a quantum number $\ell\in\mathbb{N}$. Since we want to study nonspherical perturbations, we are interested in $\ell >0$.

 We are left with a linear system of homogenous, algebraic equations for the coefficients, whose nontrivial solutions can be studied by computing the corresponding characteristic polynomial. This reduces to a fourth-order polynomial equation in the scaling exponent $\gamma$, with coefficients depending on $\ell$ and $r_+$. All the roots of the polynomial, which provide the non-trivial solutions to the homogenous equations, are real. We can eliminate two of the four solutions with boundary conditions at the horizon. Since the two \emph{smallest} roots are negative, the corresponding perturbation would blow up as $\rho \to 0$. We therefore discard them as our choice of boundary conditions. The remaining two roots give the physical scaling of the nonspherical perturbations near the $\rho = 0 $ horizon. If one of them is again negative, it cannot be removed by boundary conditions, and indicates that the perturbation is singular on the horizon. 

It is convenient to view the roots  as  functions of the dimensionless horizon radius $y_+ = r_+/L$. The largest two roots are:
\begin{subequations}
\begin{equation}
\gamma_{\pm}(\ell,y_+)=\frac{1}{2}\left[\sqrt{5-\lambda _{\ell }-12 \beta_+ \lambda _{\ell }\pm 8 \sqrt{\frac{1}{4}+12 \left(\beta_+^2-\frac{1}{144}\right) \lambda _{\ell }}}-1\right]\,,
\label{eq:lotsgammas}
\end{equation}
where
\begin{equation}
\beta_+\equiv \frac{L_{\rm AdS_2}^2}{L^2}-\frac{1}{6}\quad\text{and}\quad \lambda_{\ell}=\ell(\ell+2)\,.
\end{equation}
\end{subequations}%
Note that $\lambda_{\ell}$ is the eigenvalue of spherical harmonics on $S^3$ and $L_{\rm AdS_2}/L^2$ is only a function of $y_+$ (see  Eq.~(\ref{eq:AdS2radius})).
 
 It remains then to check whether $\gamma_{\pm}(\ell,y_+)$ are positive. It turns out that $\gamma_+(\ell,y_+)$ is positive for all values of $\ell$ and $y_+$ but $\gamma_{-}(\ell,y_+)$ is not. In particular, {\bf for $\ell = 2$, $\gamma_-$ is  negative for all $y_+ > 0$! }  Since a generic linear perturbation will always contain the $\ell = 2$ mode with some coefficient, we conclude that generic nonspherical perturbations blow up on the $AdS_2\times S^3$ horizon. Nonlinearly, even if {one} starts with a deformation on the boundary that does not include an $\ell = 2$ mode, it will be generated as one evolves in to smaller radius. This means that from the standpoint of the RG flow of a dual field theory, $AdS_2\times S^3$ is an unstable IR fixed point.
 
 We now comment on the $\ell=0,1$ modes. In this case one has to repeat the above analysis separately, since some of the structures that are used to decompose our perturbations with respect to coordinate transformations on $S^3$ turn out to vanish. A deformation with $\ell=0$ corresponds to an infinitesimal deformation of the background charge. For $\ell=1$, the calculation is more subtle. Once the dust settles, one finds a single pair of modes, with one being negative and another positive. Again, we discard the smallest exponent based on boundary conditions as $\rho\to0$. We thus restrict to the positive exponent, which is precisely given by $\gamma_+(1,y_+)$ with $\gamma_{+}(\ell,y_+)$ given in Eq.~(\ref{eq:lotsgammas}). 
   
\begin{figure}[th]
\centering \includegraphics[width=0.8\textwidth]{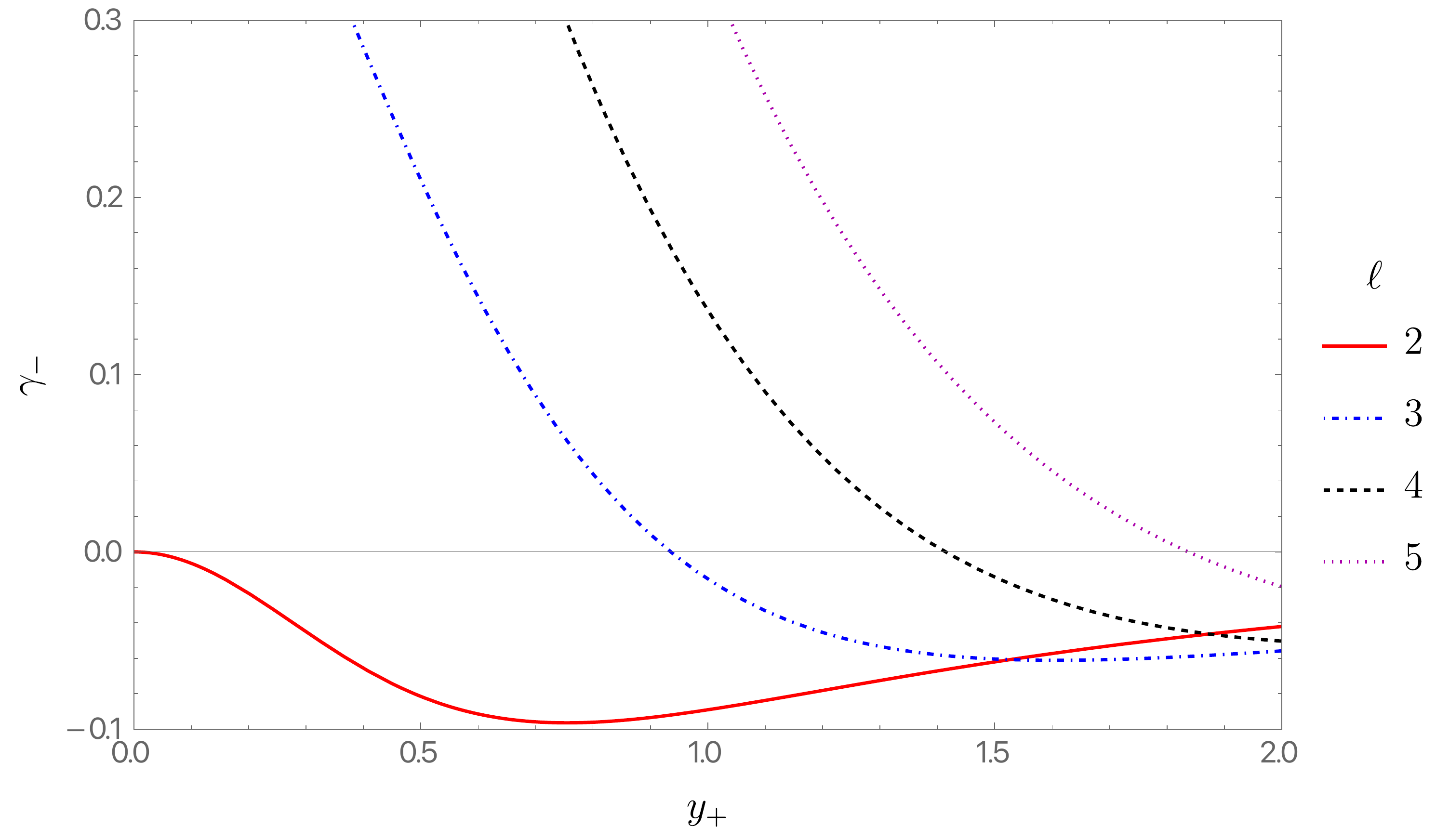}
\caption{\label{fig:gamma_negative} The scaling exponents $\gamma_-$ for perturbations of $AdS_2\times S^3$, as a function of $y_+ = r_+/L$, computed for several values of $\ell$ shown on the legend on the right.}
\end{figure}

For modes with $\ell\geq3$,  $\gamma_{-}(\ell,y_+)$ becomes negative whenever the horizon is large enough. The condition is:
 \begin{equation}
 y_+\geq y_+^c(\ell)=\frac{1}{2 \sqrt{2}}\sqrt{(\ell -2) (\ell +4)}\,,
 \label{eq:yc}
 \end{equation}
  In Fig.~\ref{fig:gamma_negative} we plot $\gamma_{-}(\ell,y_+)$  for several values of $\ell$. Note that these higher $\ell$ modes become more divergent on the horizon of a large black hole.
  
\begin{figure}[th]
\centering \includegraphics[width=0.8\textwidth]{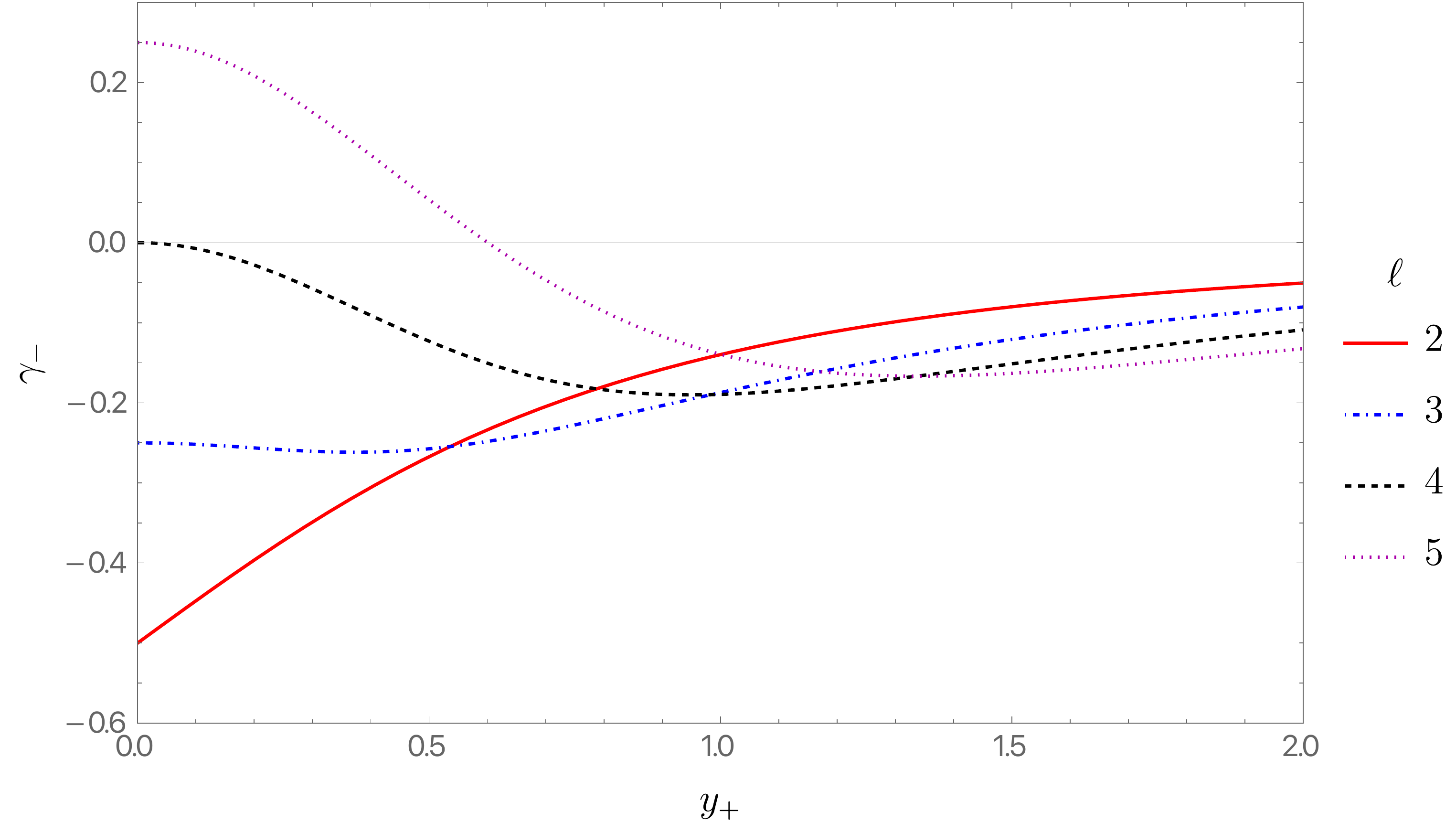}
\caption{\label{fig:gamma_negative_n} The scaling exponents $\gamma_-$ for perturbations of $AdS_2\times S^5$, as a function of $y_+ = r_+/L$, computed for several values of $\ell$ shown on the legend on the right. }
\end{figure}

The instability of the near horizon geometry that we found above in $D=5$ becomes even worse in higher dimensions, where the solution is $AdS_2\times S^n$.
The computation detailed above can be readily generalised to $n>3$. The corresponding scaling dimensions are now given by
\begin{subequations}
\begin{multline}
\gamma_{\pm}(\ell,y_+)=\frac{1}{2}\vast\{5-\frac{2\,\lambda _{\ell }}{n-1}-\frac{4\,n\,(n+1)\,\lambda _{\ell }}{(n-1)^2}\beta_+
\\
\pm\frac{4\,(n+1)}{n-1} \sqrt{4\,n\,\lambda _{\ell } \left[\beta_+
   ^2-\frac{1}{4\,n^2}\left(\frac{n-1}{n+1}\right)^2\right]+\left(\frac{n-1}{n+1}\right)^2}\vast\}^{1/2}-1\,,
\end{multline}
with
\begin{equation}
\beta_+=\frac{y_+^2}{(n-1)^2+n (n+1) y_+^2}-\frac{1}{2 n}\quad \text{and}\quad \lambda_\ell=\ell(\ell+n-1)\,.
\end{equation}
\end{subequations}
There are now more modes that are always unstable.
In analogy with the $D=5$ ($n=3$) case, we see that for $\ell=n-1$, we have
\begin{subequations}
\begin{equation}
\gamma_-(n-1,0)=0\,.
\end{equation}
However,
\begin{equation}
\gamma_-(2,0)=\frac{1}{2} \left(\frac{| n-5| }{n-1}-1\right)\,,
\end{equation}
\end{subequations}%
which is negative for all $n>3$.
In fact, for $\ell\le n-1$, $\gamma_-(\ell,y_+)<0$ for all $y_+ > 0$.  Which modes dominates near the horizon depends on the size of the black hole. This is illustrated in 
Fig.~\ref{fig:gamma_negative_n} where we plot $\gamma_{-}(\ell,y_+)$ for $n=5$ and $\ell=2,3,4,5$. Since Reissner-Nordstr\"om with $\Lambda = 0$ corresponds to $y_+ = r_+/L =0$, this shows that the near horizon region of RN in $D>5$ is also unstable to nonspherical perturbations.

\section{A new $SO(3)$-invariant IR geometry}

In this section we construct a new near horizon geometry which is only invariant under an $SO(3)$ subgroup of the $SO(4)$ rotational symmetry. We will show that it is stable to small $SO(3)$-invariant perturbations. Since charge is conserved in Einstein-Maxwell theory, we need a near horizon geometry for each $Q$. For $AdS_2 \times S^3$ this just corresponds to a trivial overall rescaling. However, our new solutions will depend nontrivially on $Q$, so we actually construct a one parameter family of new IR geometries. 

These new geometries can be written as a warped product of $AdS_2$ with a deformed three-sphere, where the $AdS_2$ length scale depends on the angles of the $S^3$. The $S^3$ is deformed in such a way that preserves a round $S^2$. In order to describe these solutions we first introduce an angle $\theta\in[0,\pi]$ and write the metric on the unit round $S^3$ as
\begin{equation}
\mathrm{d}\Omega_3^2=\mathrm{d}\theta^2+\sin^2\theta\mathrm{d}\Omega_2^2
\end{equation}
where $\mathrm{d}\Omega_2^2$ is the metric on a unit radius round $S^2$.

We then write our full IR metric and gauge field configuration as
\begin{subequations}
\begin{equation}
\mathrm{d}s^2= L^2\left\{B(\theta)\left(-A_0^2\,\rho^2\frac{\mathrm{d}t^2}{L^2}+\frac{\mathrm{d}\rho^2}{\rho^2}\right)+Y_+^2\left[H(\theta)^2\mathrm{d}\theta^2+\frac{\sin^2\theta}{H(\theta)} \mathrm{d}\Omega_2^2\right]\right\}
\label{eq:toto}
\end{equation}
and
\begin{equation}
A=-\rho_{\rm IR}\,A_0\,\rho\,\mathrm{d}t\,,
\label{eq:ansatzA}
\end{equation}
\end{subequations}
Note that the factor in parenthesis is just $AdS_2$  with the Poincare horizon at $\rho = 0$. Our \emph{Ansatz} depends on two constants, $Y_+$ and $\rho_{\rm IR}$, that determine the size of the horizon and charge density respectively. We have introduced a third constant $A_0$ that just rescales $t$. It will play no role in constructing the near horizon geometries, but will be useful when we later relate these geometries to full asymptotically AdS solutions. The function $B(\theta)$ describes the warping of the $AdS_2$, and the function $H(\theta)$ describes the distortion of the $S^3$. The entropy, $S$, and total electric charge, $Q$, of our field configuration are given by
\begin{equation}
S = \frac{\pi^2 Y_+^3}{2 G_5}\quad\text{and}\quad Q = \frac{\rho_{\rm IR}}{G_5}\,Y_+^3\,\int_0^\pi \frac{\sin^2\theta}{B(\theta)}\mathrm{d}\theta\,,
\end{equation}
Note that since charge is conserved, $Q$  can be computed at the boundary or the horizon.

The Maxwell equation is automatically satisfied using Eq.~(\ref{eq:ansatzA}), whereas the Einstein equation yields a pair of nonlinear ordinary differential equations 
\begin{subequations}
\label{eqs:nonlinear}
\begin{equation}
\left(\frac{\sin^2\theta}{H^2} B^\prime\right)^\prime+2\,Y_+^2\,\sin^2\theta\,\left(1-\frac{4}{3}\frac{\rho_{\rm IR}^2}{B}-4B\right)=0
\label{eq:first}
\end{equation}
where $^\prime$ denotes differentiation with respect to $\theta$ and
\begin{equation}
\frac{1}{B^2\sin^2\theta}\left[\left(\frac{B^2 \sin^2\theta}{H}\right)^\prime\right]^{2}-4 B^2 H-\frac{3 \sin^2\theta}{H^2}{B^\prime}^2-4\,Y_+^2\,\sin^2\theta\left(\rho_{\rm IR}^2-B+6 B^2\right)=0\,.
\end{equation}
\label{eqs:finalIR}
\end{subequations}%
Note that, as advertised, $A_0$ does not appear in these equations of motion. Regularity at the poles requires the boundary conditions $B^\prime(0)=B^\prime(\pi)=0$ and $H(0)=H(\pi)=1$.

From Eq.~(\ref{eq:first}) it is clear that $\rho_{\rm IR}$ is not a free parameter. Indeed, one can integrate both sides of Eq.~(\ref{eq:first}) and use the above boundary conditions to find the following relation
\begin{equation}
\int_0^\pi\,\sin^2\theta\,\left[1-\frac{4}{3}\frac{\rho_{\rm IR}^2}{B(\theta)}-4B(\theta)\right]\mathrm{d}\theta=0\,,
\label{eq:globalcon}
\end{equation}
Thus, although  our \emph{Ansatz} depends on two free parameters $(Y_+,\rho_{\rm IR})$ the relation above fixes one of them, so that we only have a single parameter free.

\subsection{Perturbative analytic treatment}\label{sec:ell2}
We have not managed to find closed form solutions of Eqs.~(\ref{eqs:finalIR}). We have, however, found an analytic perturbative scheme which we can extend to whatever order in perturbation theory we wish. The idea is the following.  We have seen in section \ref{sec:unstable} that $\ell=2$ perturbations have $\gamma_-(2,0)=0$. This suggests that there might exist a new family of near horizon geometries with $AdS_2$ symmetry that branches off from the zero size limit of $AdS_2\times S^2$. Since the zero size limit  is singular, this is an unconventional perturbation expansion. Nevertheless we will see that it is well defined.

 We thus expand
\begin{subequations}
\begin{align}
&B(\theta)=\sum_{i=1}^{+\infty}b^{(i)}(\theta)\varepsilon^i\,,
\\
&H(\theta)=1+\sum_{i=1}^{+\infty}h^{(i)}(\theta)\varepsilon^i\,,
\\
& \rho_{\rm IR}^2=\sum_{i=1}^{+\infty}\Xi^{(i)}\varepsilon^i\,,
\\
& Y_+^2=\sum_{i=1}^{+\infty}\Sigma_{+}^{(i)}\varepsilon^i\,,
\end{align}
\label{eq:per}
\end{subequations}%
where $\varepsilon$ is a book-keeping parameter, whose normalisation we choose to be
\begin{equation}
\int_0^{\pi}B \sin^2\theta\,Y_{\ell=2}(\theta) \mathrm{d}\theta=\varepsilon^2\,.
\end{equation}
where $Y_{\ell}(\theta)$ is a spherical harmonic on the three-sphere preserving $SO(3)$, which we choose to be given by
\begin{equation}
Y_{\ell}(\theta)=\sqrt{\frac{2}{\pi }} \frac{\sin [(\ell +1) \theta]}{\sin \theta}\quad \text{so that} \quad \int_0^{\pi}Y_{\ell}(\theta)Y_{\tilde{\ell}}(\theta)\mathrm{d}\theta=\delta_{\ell\,,\tilde{\ell}}\quad\text{and}\quad Y_{\ell}(0)>0\,.
\end{equation}
Note that $H$ starts with $1$ to satisfy our boundary conditions, and $\epsilon >0$ is required since $Y_+^2$ begins at order $\epsilon $.

Despite the fact that $\varepsilon=0$ corresponds to a \emph{singular} solution, at each order in
 $\varepsilon$ our boundary conditions are sufficient to solve for each of the functions above. In particular, for any finite value of $\varepsilon$ our perturbative expansion yields a completely smooth solution. We carried this expansion all the way to $\mathcal{O}(\varepsilon^7)$. The first few coefficients are
\begin{align}
&b^{(1)}(\theta)=\frac{\sqrt{7}}{2 (2 \pi)^{1/4}}\,,\quad b^{(2)}(\theta)=\frac{56 \cos (2 \theta )-61}{14 \sqrt{2 \pi }}\,, \nonumber
\\
& b^{(3)}(\theta)=\frac{588 \cos (2 \theta )+588 \cos (4 \theta )+6354}{147 \sqrt{7} (2 \pi )^{3/4}}\,,\nonumber
\\
&h^{(1)}(\theta)=-\frac{4\ 2^{3/4} \sin ^2\theta}{\sqrt{7} \pi ^{1/4}}\,,\quad h^{(2)}(\theta)=-\frac{496}{49} \sqrt{\frac{2}{\pi }} \sin ^2\theta\,,\nonumber
\\
&h^{(3)}(\theta) = -\frac{16 2^{1/4} \sin ^2\theta}{1029 \sqrt{7} \pi ^{3/4}}\left[2891 \cos (2 \theta )+11442\right]\,,\nonumber
\\
& \Xi^{(1)}=\frac{3 \sqrt{7}}{8 (2 \pi)^{1/4}}\,,\quad \Xi^{(2)}=-\frac{561}{56 \sqrt{2 \pi }}\,,\quad \Xi^{(3)}=\frac{15525}{98 \sqrt{7} (2 \pi )^{3/4}}\,,\nonumber
\\
& \Sigma_+^{(1)}=\frac{2^{3/4} \sqrt{7}}{\pi^{1/4}}\,,\quad \Sigma_+^{(2)}=\frac{205 }{7}\sqrt{\frac{2}{\pi }}\,,\quad \Sigma_+^{(3)}=\frac{37360 2^{1/4}}{49 \sqrt{7} \pi ^{3/4}}\,.
\end{align}
Note that since the solution starts with an $\ell = 2$ perturbation, there is a reflection symmetry about $\theta = \pi/2$ which is preserved to all orders.

With the above it is a simple exercise to compute the total charge and entropy as a function of $\varepsilon$. These turn out to be given by
\begin{subequations}
\begin{multline}
\frac{G_5\,Q}{L^2}=\frac{\sqrt{21} \pi ^{3/4}}{2^{1/4}}  \varepsilon\Bigg[1+\frac{303}{7 \sqrt{7} (2 \pi)^{1/4}} \varepsilon +\frac{36369}{343} \sqrt{\frac{2}{\pi }} \varepsilon ^2+\frac{15631151}{2401 \sqrt{7} (2 \pi
   )^{3/4}}\varepsilon^3
   \\
   +\frac{682434694}{50421 \pi }\varepsilon^4+\frac{630372065550}{823543 \sqrt{7} (2 \pi )^{5/4}} \varepsilon^5+\mathcal{O}(\varepsilon^6)\Bigg]
   \label{eq:Qepsilon}
\end{multline}
and
\begin{multline}
\frac{G_5\,S}{L^3}=2^{1/8} 7^{3/4} \pi ^{13/8} \varepsilon ^{3/2} \Bigg[1+\frac{615}{14 \sqrt{7} (2 \pi )^{1/4}} \varepsilon +\frac{574395}{2744 \sqrt{2 \pi }} \varepsilon ^2+\frac{34688917}{5488 \sqrt{7} (2 \pi)^{3/4}} \varepsilon^3
   \\
   +\frac{392491070291}{30118144 \pi }\varepsilon^4+\frac{309058579837106}{421654016 \sqrt{7} (2 \pi )^{5/4}} \varepsilon ^5+\mathcal{O}(\varepsilon^6)\Bigg]\,,
\end{multline}
\end{subequations}
respectively. The parameter $\varepsilon$, though very useful for practical implementations, has little physical meaning. We shall see that the entropy of this novel solution is not very different from that of an extremal RN-AdS black hole with the same total charge $Q$. For this reason we define
\begin{equation}
\Delta S = S(Q)-S_{\rm RN}(Q)
\end{equation}
which gives the difference in entropy between one of the solutions we are seeking to construct and an extreme RN-AdS black hole with the same total charge $Q$. It is then a simple exercise to compute $\Delta S$ as a function of $\varepsilon$ (or alternatively, $Q$ through Eq.~(\ref{eq:Qepsilon})). The final result, consistent with our $\mathcal{O}(\varepsilon^7)$ expansion for $B$ and $H$, turns out to be

\begin{align}
\frac{G_5\,\Delta S}{L^3} & = -2^{3/8} 7^{1/4} 8\pi ^{7/8} \varepsilon^{9/2}-\frac{3468 2^{1/8} \pi ^{5/8}}{7^{5/4}} \varepsilon^{11/2}-\frac{987123 \pi ^{3/8}}{2^{1/8} 7^{3/4}49 } \varepsilon^{13/2}+\mathcal{O}(\varepsilon^{15/2})\nonumber
\\
& =-\frac{16 \sqrt{2}}{441 \pi ^{5/2}3^{1/4}} \left(\frac{G_5\,Q}{L^2}\right)^{9/2} \Bigg\{1-\frac{310 \sqrt{3}}{49 \pi}\left(\frac{G_5\,Q}{L^2}\right)+\frac{415279}{4802 \pi ^2} \left(\frac{G_5\,Q}{L^2}\right)^2\nonumber
\\
&\qquad\qquad\qquad\qquad\qquad\qquad\qquad\qquad\qquad\qquad+\mathcal{O}\left[\left(\frac{G_5\,Q}{L^2}\right)^3\right]\Bigg\}
\end{align}
This analytic expression works remarkably well when $Q/L^2\ll1$.

\subsection{\label{sec:exact}Exact numerical results}
We now solve Eqs.~(\ref{eqs:nonlinear}) fully non-linearly using numerical methods. It is easy to see that, at least locally, Eqs.~(\ref{eqs:nonlinear}) gives a one-parameter family of solutions. It might appear that Eqs.~(\ref{eqs:nonlinear}) depends on two parameters, $\rho^2_{\rm IR}$ and $Y_+^2$, but because of the global constraint in Eq.~(\ref{eq:globalcon}), one of these parameters gets locked in terms of the other.

We discretize the $\theta$ direction with a Chebyshev-Gauss-Lobatto collocation grid, and solve the resulting equations using a Newton-Raphson method. These methods have been reviewed in the literature in \cite{Dias:2015nua}. The size of the horizon is determined by $Y_+$, and the charge $Q$ is a monotonically increasing function of $Y_+$.  We are able to construct solutions for all $Q$ up to about $G_5 Q/L^2\sim 3\times 10^{4}$ without encountering any numerical issues. We believe they extend to arbitrarily large $Q$.

\begin{figure}[th]
\centering \includegraphics[width=0.75\textwidth]{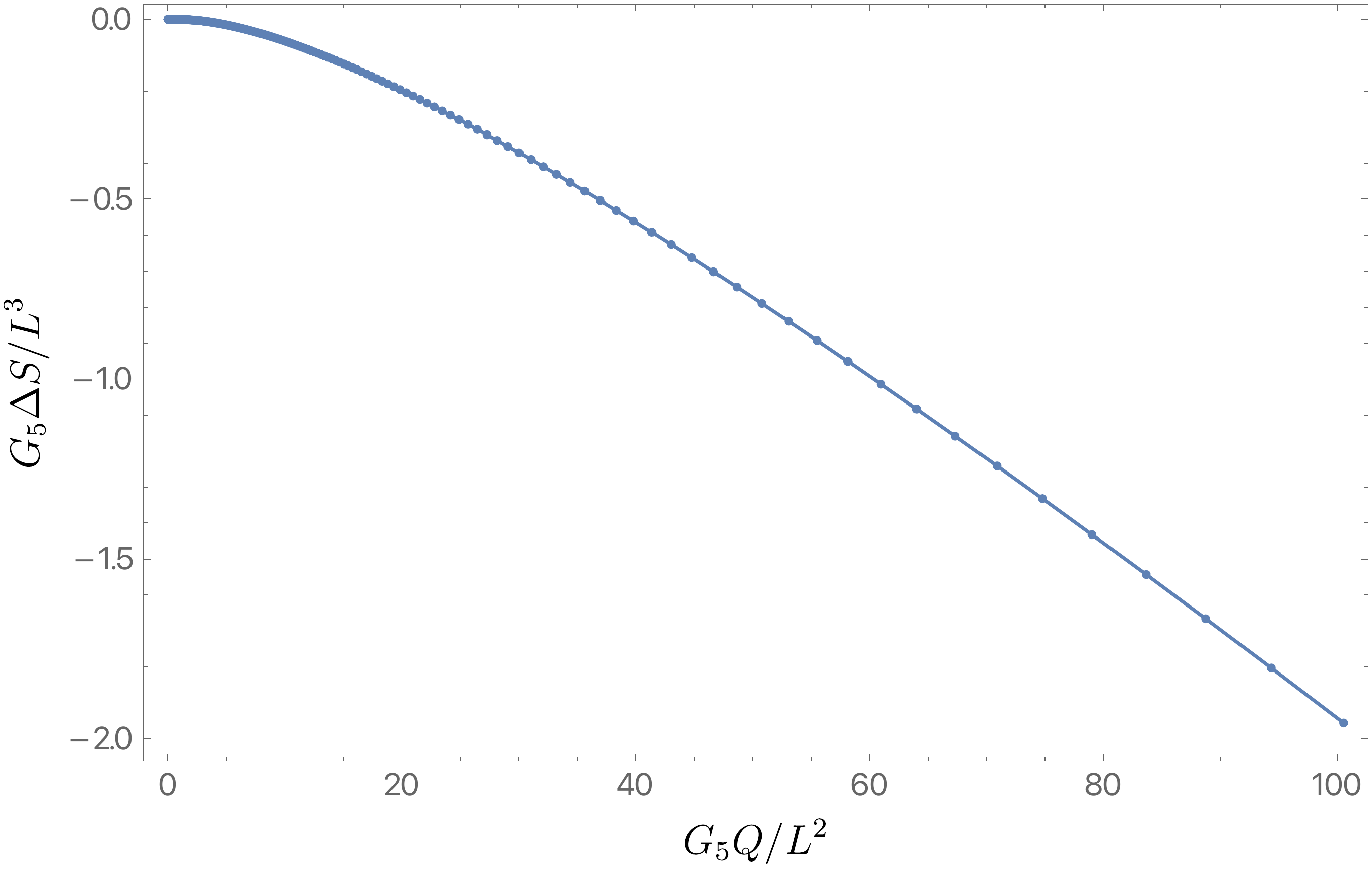}
\caption{\label{fig:large} The difference in entropy between the new near horizon geometries and RN AdS, as a function of charge.}
\end{figure}

 In Fig.~\ref{fig:large} we show $\Delta S$ for charges up to $G_5 Q/L^2\sim 100$. Note that $\Delta S < 0$ for all $Q$, showing that the new IR geometries have smaller entropy than RN AdS with the same charge.

\begin{figure}[th]
\centering \includegraphics[width=0.4\textwidth]{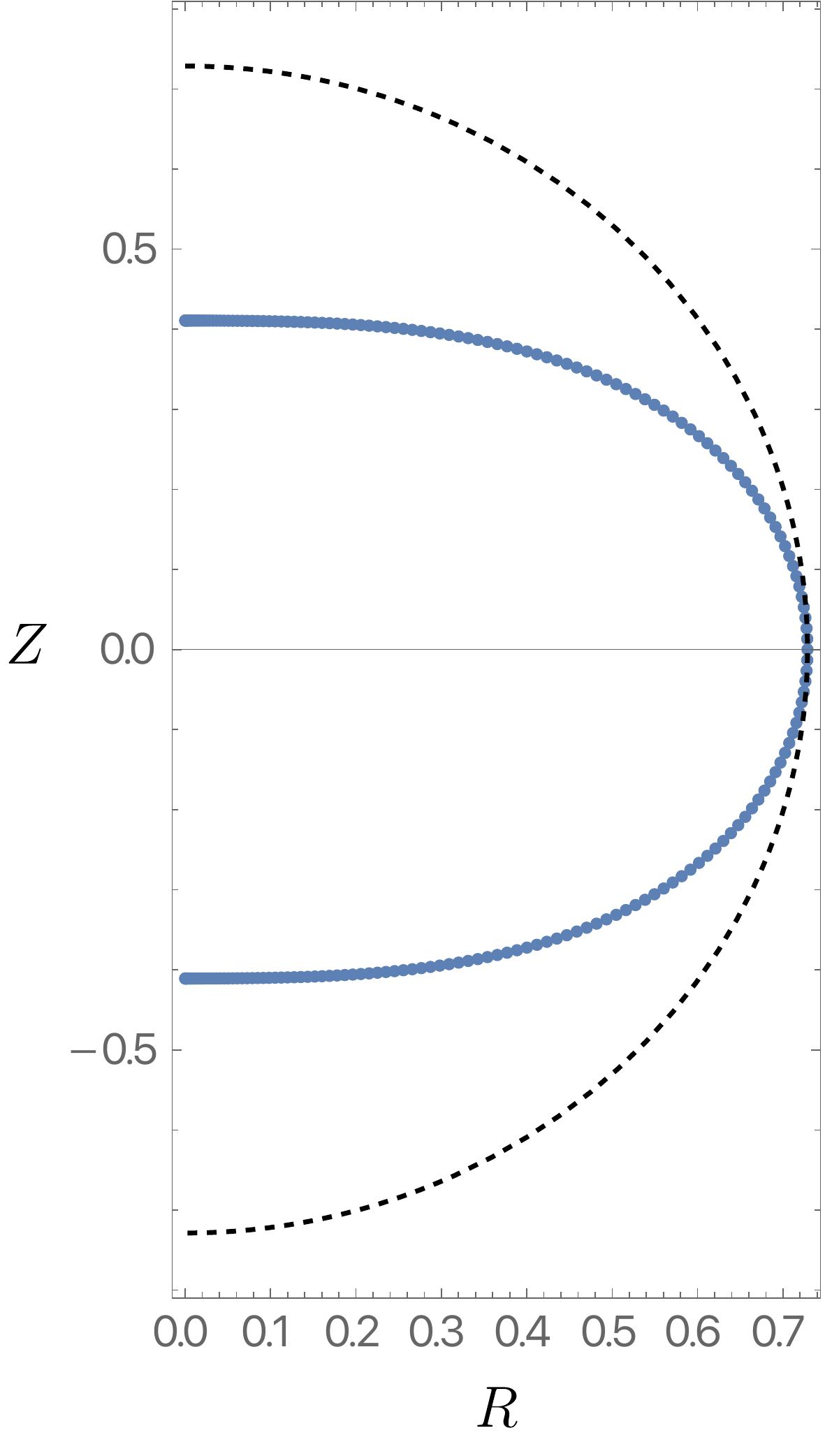}
\caption{\label{fig:embedding} Isometric embedding of the near horizon geometry into $\mathbb{R}^4$. The black dashed line shows what a perfect sphere would look like and the blue disks represent our novel pancaked IR geometry. This particular embedding was generated for $G_5 Q/L^2\approx 1.61$.}
\end{figure}

We now explore the geometry of a spatial cross section of the horizon. It is topologically $S^3$, but  is no longer round. The reflection symmetry about  $\theta = \pi/2$ that we saw in the perturbative solution  remains in the exact solutions for all $Q$.
To begin, let us try to embed it into $\mathbb{R}^4$. To do this we set
\begin{align}
&y_1=L\,Z(\theta)\,, \nonumber
\\
&y_2=L\,R(\theta)\cos \theta_1\,, \nonumber
\\
&y_3=L\,R(\theta)\sin \theta_1\cos \phi\,, \nonumber
\\
&y_4=L\,R(\theta)\sin \theta_1\sin \phi\,, \nonumber
\end{align}
with $\theta_1\in[0,\pi]$ and $\phi\sim\phi+2\pi$ the usual latitude and longitude angles on a two-sphere, respectively. We then compare the induced metric on
\begin{equation}
\mathrm{d}s^2=\sum_{i=1}^4\mathrm{d}y_i^2
\end{equation}
with that of a spatial cross section of our horizon obtained from  Eq.~(\ref{eq:toto}). We thus obtain
\begin{equation}
R(\theta)=\frac{Y_+ \sin\theta}{\sqrt{H(\theta)}}\quad\text{and}\quad Z^\prime(\theta)^2=Y_+^2 H(\theta)^2-R^\prime(\theta)^2\,.
\label{eq:emb}
\end{equation}
The latter equation can be solved using numerical methods.  For small $Y_+$ (i.e. small $Q$) the horizon is only slightly distorted from a round $S^3$, consistent with the previous perturbative results.  For sufficiently large values of $Y_+$ (i.e. large $Q$) there is no solution to \eqref{eq:emb}, showing that the near horizon geometry stops being embeddable into $\mathbb{R}^4$.  This is similar to Kerr and Kerr-Newman black holes near extremality  \cite{Smarr:1973zz}. The isometric embedding of the horizon for $G_5 Q/L^2 \approx 1.61$ is shown in Fig.~\ref{fig:embedding}. One can see that the horizon becomes flattened like a pancake. The black dashed line shows what a perfect sphere would look like, for comparison. The blue disks correspond to the numerical embedding obtained by solving Eq.~(\ref{eq:emb}). 

\begin{figure}[th]
\centering \includegraphics[width=0.9\textwidth]{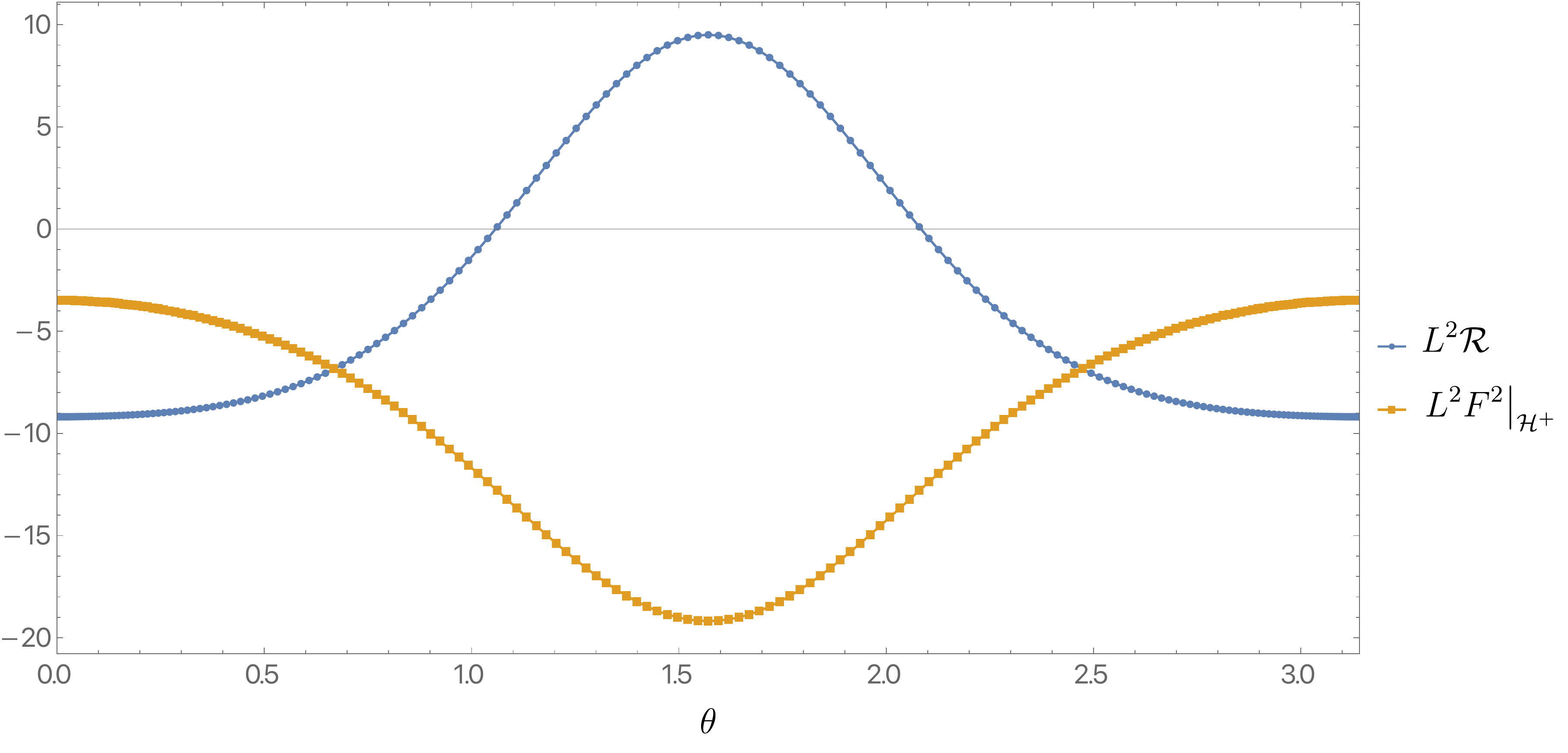}
\caption{\label{fig:ricci_local} $\mathcal{R}$ and $F^2$ on the horizon, as a function of $\theta$ for $Q\approx 100.523$.}
\end{figure}

 In order to picture the large $Q$ solution, we first plot  $\mathcal{R}$ and $F^2$ on the horizon, as a function of $\theta$ for $G_5 Q /L^2 \approx 100$. This is shown in Fig.~\ref{fig:ricci_local}.
We can see that near the equator, \emph{i.e.} $\theta=\pi/2$, the Ricci scalar $\mathcal{R}$ is positive as expected, but $\mathcal{R}$ becomes {\it negative} near the poles. In addition, we see that the electric field is stronger at the equator and weaker near the poles.

\begin{figure}[th]
\centering \includegraphics[width=0.9\textwidth]{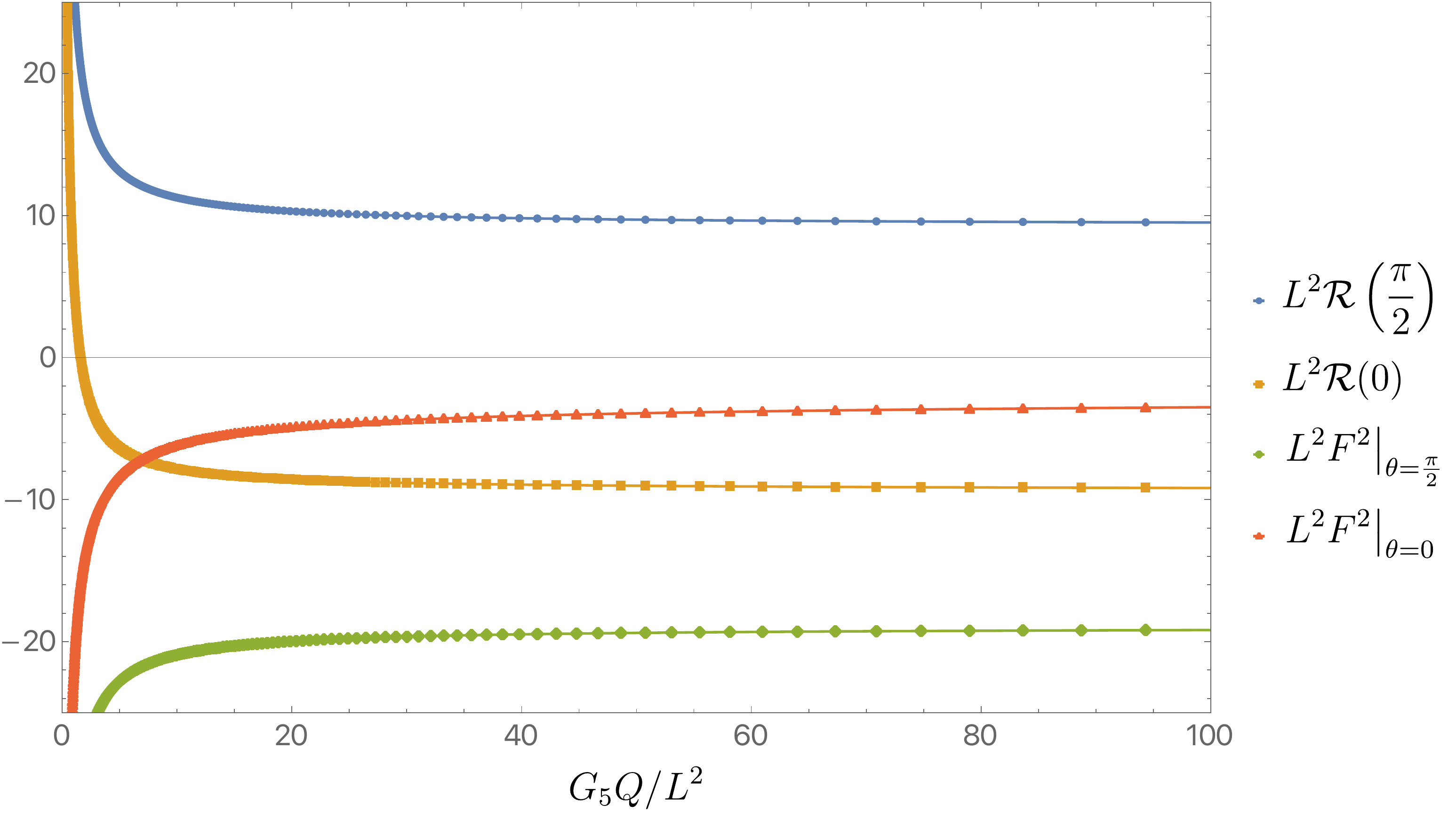}
\caption{\label{fig:curvature_poles_equator} $\mathcal{R}(\pi/2)$ (blue disks), $\mathcal{R}(0)$ (orange squares), $\left.F^2\right|_{\theta=\frac{\pi}{2}}$ (green diamonds) and $\left.F^2\right|_{\theta=0}$ (red triangles) as a function of $Q$.}
\end{figure}

To map out how a round sphere with uniform electric field (for small $Q$) evolves to something like Fig.~\ref{fig:ricci_local} (for large $Q$), we plot
 $\mathcal{R}(0)$, $\mathcal{R}(\pi/2)$, $\left.F^2\right|_{\theta=0}$ and $\left.F^2\right|_{\theta=\frac{\pi}{2}}$ as one increases $Q$. This is shown in Fig.~\ref{fig:curvature_poles_equator}. One clearly sees that the curvature at the poles decreases rapidly as $Q$  increases from the large positive curvature of a small sphere to a constant negative value. The curvature at the equator also decreases but settles down to a constant positive value.
 The limiting behavior at large $Q$ is
\begin{subequations}
\begin{equation}
\lim_{Q\to+\infty}L^2\mathcal{R}(0)\equiv \mathcal{R}_0 \approx -9.3913\,,\qquad\lim_{Q\to+\infty}L^2\mathcal{R}(\pi/2)\equiv \mathcal{R}_e\approx 9.3058\,,
\label{eq:limitingscalars}
\end{equation}
and
\begin{equation}
\lim_{Q\to+\infty}\left.F^2\right|_{\theta=0} \approx -2.5947\,,\qquad \lim_{Q\to+\infty}\left.F^2\right|_{\theta=\frac{\pi}{2}}\approx -18.9797\,.
\label{eq:limitingscalarsF}
\end{equation}
\end{subequations}
Since the horizon volume is growing with $Q$, but the curvature is not decreasing, the horizon must look like a large three-dimensional hyperbolic space near each pole, joined together by a positive curvature ring around the equator.

To understand the limiting geometry more explicitly, we will change gauge. Consider
\begin{equation}
\mathrm{d}s^2= L^2\left\{B(\chi)\left(-A_0^2\,\rho^2\frac{\mathrm{d}t^2}{L^2}+\frac{\mathrm{d}\rho^2}{\rho^2}\right)+\tilde{Y}_+^2\,\left[\tilde{H}(\chi)\mathrm{d}\chi^2+\sin^2\chi\, \mathrm{d}\Omega_2^2\right]\right\}\,
\end{equation}
instead of Eq.~(\ref{eq:toto}). This amounts to a simple change of coordinates. We are interested in the large $\tilde{Y}_+$ limit of the resulting equations of motion. The advantage of this coordinate system is that we have to solve just a single second order equation of motion for $B$. Indeed, after some algebra, we find that $\tilde H$ can be expressed in terms of $B$ and its first derivative:
\begin{subequations}
\begin{equation}
\tilde{H}(\chi)=\frac{1}{4}\frac{1 }{B^2+\tilde{Y}_+^2 \left[\rho_{\rm IR}^2-B\,(1-6 B)\right] \sin ^2\chi}\left[\left(\sin\chi\,\frac{\partial B}{\partial \chi}+4 \cos\chi\,B\right)^2-12\,B^2\,\cos^2\chi\right]\,.
\label{eq:algebraic}
\end{equation}
while $B(\chi)$ satisfies the following second order differential equation
\begin{equation}
\frac{\partial }{\partial \chi }\left(\frac{\sin ^2\chi}{\sqrt{\tilde{H}}}\frac{\partial B}{\partial \chi }\right)-\frac{2 \tilde{Y}_+^2 \sqrt{\tilde{H}}}{3 B} \sin ^2\chi \left(4 \rho_{\rm IR}^2-3 B+12 B^2\right)=0\,.
\label{eq:singB}
\end{equation}
\end{subequations}
Note that near the poles, located at $\chi=0\,,\pi$, Eq.~(\ref{eq:algebraic}) automatically yields $\tilde{H}=1$, as it should from regularity. The solution we seek to construct is even around $\chi=\pi/2$, so we have
\begin{equation}
\left.\frac{\partial B}{\partial \chi}\right|_{\chi=\frac{\pi}{2}}=0\,.
\end{equation}
From Eq.~(\ref{eq:algebraic}) we find that in order for $H(\pi/2)$ \emph{not} to vanish we must demand
\begin{equation}
\rho^2_{\rm IR}=B\left(\frac{\pi }{2}\right) \left[1-B\left(\frac{\pi }{2}\right) \left(6+\frac{1}{\tilde{Y}_+^2}\right)\right]\,.
\end{equation}
The above equation, so long as $B(\pi/2)$ is non-vanishing, provides a simple relation between $\rho^2_{\rm IR}$ and $B(\pi/2)$. In particular, in the large $\tilde{Y}_+$ limit we find
\begin{equation}
\rho^2_{\rm IR}=B\left(\frac{\pi }{2}\right) \left[1-6\,B\left(\frac{\pi }{2}\right)\right]\,.
\label{eq:B2}
\end{equation}

We would like to find a similar relation, in the large $\tilde{Y}_+$ limit, between $\rho^2_{\rm IR}$ and $B(0)$. However, it is clear from Eq.~(\ref{eq:algebraic}) that the expansion near $\chi=0$ needs to be treated with care, since the factor of $\tilde{Y}_+^2$ appearing in the denominator comes multiplied by $\sin^2\chi$. In order to deal with this, we first change into a new variable 
\begin{equation}
\xi \equiv \tilde{Y}_+^2\sin \chi
\end{equation}
and take the large $\tilde{Y}_+$ limit \emph{a posteriori}, while keeping $\xi$ fixed. This ensures that while we take $\tilde{Y}_+$ to be large, we are zooming in to $\chi=0$. This procedures then yields
\begin{equation}
\rho_{\rm IR}^2=\frac{3}{4} \left[1-4 B(0)\right] B(0)\,,
\label{eq:B1}
\end{equation}
to leading order at large $\tilde{Y}_+$. We have thus found a relation between $B(0)$ and $B(\pi/2)$ for large horizons, by combining Eq.~(\ref{eq:B2}) and Eq.~(\ref{eq:B1}).

We can use this result to obtain analytic expressions relating  $\mathcal{R}$ and $F^2$ at the equator and at the poles.
 Let us start with $\mathcal{R}$. This is a function of $\tilde{H}(\chi)$ and its first derivative only. However, from Eq.~(\ref{eq:algebraic}) we can alternatively express $\mathcal{R}$ as a function of $B$ and its first and second derivatives. By using the equation of motion for $B$ (see Eq.~\ref{eq:singB}) we can eliminate all second derivatives, thus finding an expression for $\mathcal{R}$ as a function of $B$ and its first derivative only. Finally we note that the first derivative of $B$ vanishes at $\theta=0,\pi/2$, so we are left with an expression for $\mathcal{R}(0)$ and $\mathcal{R}(\pi/2)$ as a function of $B(0)$ and $B(\pi/2)$, respectively.

We can now substitute in the above relation between $B(0)$ and $B(\pi/2)$ to obtain an analytic relation between $ \mathcal{R}_0 \equiv \mathcal{R}(0)$ and  $\mathcal{R}_e \equiv\mathcal{R}(\pi/2)$, valid for large $Q$. The result is:

\begin{equation}
\mathcal{R}_0 = \frac{9}{128 (\mathcal{R}_e+16)} \left[\left(3 \mathcal{R}_e+32\right) \left(\mathcal{R}_e-32\right)-\left(\mathcal{R}_e+32\right) \sqrt{9 \mathcal{R}_e^2+64 \mathcal{R}_e+1024}\right]\,.
\end{equation}
We have tested this relation with the values in Eq.~(\ref{eq:limitingscalars}) and find that it matches the numerical results to within $0.1\%$. 

Since $F^2$ only depends on $B$ and $\rho_{IR}$ (which is determined in terms of $B$ via \eqref{eq:B2} or \eqref{eq:B1})
we clearly have a relation between $F^2$ at $\chi=0$ and at $\chi =\pi/2$. But we can also express both of them in terms of $\mathcal{R}_e$:
\begin{subequations}
\begin{equation}
\lim_{\tilde{Y}_+\to+\infty}\left.L^2 F^2\right|_{\chi=\frac{\pi}{2}}=-\frac{3}{4} \left(\mathcal{R}_e+16\right)\,,
\end{equation}
and
\begin{equation}
\lim_{\tilde{Y}_+\to+\infty}\left.L^2 F^2\right|_{\chi=0}=-\frac{3072 \left(\mathcal{R}_e+16\right)}{\left(3 \mathcal{R}_e+96+\sqrt{9 \mathcal{R}_e^2+64 \mathcal{R}_e+1024}\right)^2}\,.
\end{equation}
\end{subequations}
Using the value quoted in Eq.~(\ref{eq:limitingscalars}) for $\mathcal{R}_e$, we find that these expressions reproduce the values quoted in Eq.~(\ref{eq:limitingscalarsF}) to within $0.25\%$. So we see that the large $Q$ limits of the curvature and Maxwell field at the equator and the pole are all determined by $\mathcal{R}_e$.
\subsection{RG stability of the new IR geometries with respect to $SO(3)$ preserving deformations}\label{RGstable}
In this section we study the RG stability of our new near horizon geometries. The analysis developed here has a drawback: it is a linear analysis and it could well be that nonlinearities change the overall picture. In the next section we study fully nonlinear deformations and show that this is not the case.

Before proceeding let us briefly discuss what the expectations are. When we studied the RG stability of $AdS_2\times S^3$, we decomposed all perturbations in terms of spherical harmonics on $S^3$, which in turn were labelled by a quantum number $\ell$. For  each value of $\ell$ we can find a total of \emph{four} scaling exponents $\gamma$. Two of which are eliminated via boundary conditions at the horizon, since they turn out to always be negative. The remaining two roots are then studied as a function of $y_+$, or equivalently $Q$. We would like to keep this procedure as much as possible.

However, once we break $SO(4)$ we need to find a way to articulate what we mean by a perturbation having a certain $\ell$. We do this by counting nodes along the $\theta$ direction. This allows us to make sense of $\ell$ beyond spherical symmetry. Note that a given standard $\ell$-harmonic on the three-sphere \emph{does} have $\ell$ nodes along the polar direction. For each value of $\ell\neq 0,1$ we are then supposed to find four values for the scaling exponents. We discard the two most negative exponents, which connect to the unphysical scaling exponents when $SO(4)$ symmetry is restored (\emph{i.e.} $Q=0$). Unlike for the perturbations of  $AdS_2\times S^3$, we now need to resort to solving an honest quadratic St\"urm-Liouville problem, which we will detail next.

First, we  present our perturbative Ansatz, which is a function of the scaling exponents $\gamma$. We then take $g=\bar{g}+h$, $A=\bar{A}+a$,  with bared quantities being our novel IR geometries, and set
\begin{subequations}
\begin{multline}
\delta \mathrm{d}s^2\equiv h_{ab}\mathrm{d}x^a\mathrm{d}x^b= \rho^\gamma\,L^2\Bigg\{B(\theta) \frac{q_1(\theta)}{\gamma}\,\left(-A_0^2\,\rho^2\frac{\mathrm{d}t^2}{L^2}+\frac{\mathrm{d}\rho^2}{\rho^2}\right)
\\
+Y_+^2\left[q_2(\theta)\,H(\theta)^2\mathrm{d}\theta^2+q_3(\theta)\,\frac{\sin^2\theta}{H(\theta)} \mathrm{d}\Omega_2^2\right]\Bigg\}
\label{eq:petoto}
\end{multline}
and
\begin{equation}
\delta A\equiv a_a \mathrm{d}x^a=-\rho_{\rm IR}\,A_0\,\rho^{1+\gamma}\,q_4(\theta)\,\mathrm{d}t\,.
\label{eq:peansatzA}
\end{equation}
\end{subequations}
Note that the perturbations preserve an $SO(3)$ symmetry. 
This form of the metric is already gauged fixed, in the sense that $h_{tt}$ and $h_{\rho\rho}$ are not independent components, and metric components of the form $h_{\rho \theta}$ are absent. These conditions fix both infinitesimal reparametrisations of $\theta$ and $\rho$, as it should. It then remains to find $q_1$, $q_2$, $q_3$, $q_4$ and $\gamma$ from the Einstein-Maxwell equations.

The procedure is somehow tedious, so we will only present the final results.  Setting  $\lambda = \gamma(\gamma+1)$, we find
\begin{subequations}
\begin{equation}
q_2=-2\,q_3\,,
\label{eq:deltaq2}
\end{equation}
\begin{multline}
q_3=\frac{1}{\mathcal{H}_\lambda}\Bigg\{2 Y_+^2 \left[4 \rho _{\text{IR}}^2+(\lambda -2) B\right] \sin \theta  H^3 \frac{q_1}{\gamma}-8 \rho _{\text{IR}}^2 Y_+^2 (\gamma +1) \sin \theta H^3 q_4
\\
+3 B^2\left(2 \cos \theta  H-\sin \theta  H'\right) \frac{q_1'}{\gamma}- \left[4 B \sin \theta  H'-8 H\left(B \cos \theta +\sin\theta B'\right)\right]\frac{\rho _{\text{IR}}^2 q_4'}{\gamma }\Bigg\}\,,
\label{eq:deltaq3}
\end{multline}
\begin{multline}
\frac{\mathcal{H}_\lambda^2}{B^2 \sin ^3\theta H}\left(\frac{B^2 \sin ^3\theta H\,q_1'}{\mathcal{H}_\lambda}\right)^\prime+\left(\alpha_0\,\mathcal{H}_\lambda+\alpha_1+\alpha_2\,\lambda\right)q_4^\prime\\
+\left(\beta_0\,\mathcal{H}_\lambda+\beta_1+\beta_2\,\lambda+\beta_3 \lambda^2\right)q_1+\left(\kappa_0\,\mathcal{H}_\lambda+\kappa_1\,+\kappa_2 \lambda\right)\,\lambda\,q_4=0\,.
\label{eq:deltaq1}
\end{multline}
\begin{equation}
\left(\frac{\sin ^2\theta\,q_4'}{H^2}\right)^\prime-\frac{Y_+^2 \sin ^2\theta}{B} \left(  q_1-\lambda  q_4\right)=0\,.
\label{eq:deltaq4}
\end{equation}
where $\alpha_0$, $\alpha_1$, $\alpha_2$, $\beta_0$, $\beta_1$, $\beta_2$, $\beta_3$, $\kappa_0$, $\kappa_1$ and $\kappa_2$ are functions of $B$, $B^\prime$, $H$, $H^\prime$ and $\theta$ given in appendix \ref{app:crazy_long} and are independent of $\lambda$, and
\begin{equation}
\mathcal{H}_\lambda\equiv \sin \theta \Big\{4 Y_+^2 \left[\rho _{\text{IR}}^2-(\lambda +1-6 B) B\right] H^3-3 H {B^\prime}^2
+6 B B' H'\Big\}-12 B\cos\theta H  B'\,.
\end{equation}
\end{subequations}%
 Once $q_1$ and $q_4$ are known from Eq.~(\ref{eq:deltaq1}) and Eq.~(\ref{eq:deltaq4}), $q_2$ and $q_3$ are fixed in terms of Eq.~(\ref{eq:deltaq2}) and Eq.~(\ref{eq:deltaq3}). We are thus left with solving Eq.~(\ref{eq:deltaq1}) and Eq.~(\ref{eq:deltaq4}), which should determine $q_1$, $q_4$ and $\lambda$. As boundary conditions we demand
\begin{equation}
q_1^\prime(0)=q_1^\prime(\pi)=q_4^\prime(0)=q_4^\prime(\pi)=0\,,
\end{equation}
which render Eq.~(\ref{eq:deltaq1}) and Eq.~(\ref{eq:deltaq4})  a quadratic St\"urm-Liouville eigenvalue problem in $\lambda$. Note that once a solution for $q_1$ and $q_4$ has been found, we still need to a posteriori check that $q_3$ given in Eq.~(\ref{eq:deltaq3}) is everywhere smooth. The present work only discusses modes for which all of the functions $q_i$ are smooth for $\theta\in[0,\pi]$.

We solved Eq.~(\ref{eq:deltaq1}) and Eq.~(\ref{eq:deltaq4}) in two different manners, which agree well with each other in the regime where both methods are applicable. First, by using our perturbative scheme, we determine $\lambda$ as a function of $\varepsilon$ (see section \ref{sec:ell2}), which gives an expansion valid at small $Q$.
We set
\begin{subequations}
\begin{align}
&q_1(\theta) = \sum_{i=0}^{+\infty}q_1^{(i)}(\theta)\varepsilon^i
\\
&q_4(\theta) = \sum_{i=0}^{+\infty}q_4^{(i)}(\theta)\varepsilon^i
\\
&\gamma = \sum_{i=0}^{+\infty}\gamma^{(i)}\varepsilon^i
\end{align}
\end{subequations}%
and solve order by order in $\varepsilon$. The most problematic mode is the mode that goes negative for all $Q>0$ for $AdS_2\times S^3$. This mode has $\ell=2$, which is the mode we would like to disentangle.

To start our perturbative treatment we take
\begin{equation}
\gamma^{(0)}=0\,,\quad q_1^{(0)}=\sqrt{\frac{2}{\pi}}[2+\cos( 2\theta)]\qquad\text{and}\quad q_4^{(0)}=0\,.
\end{equation}
Note that $q_1^{(0)}(\theta)$ is an $\ell=2$ harmonic on a round three-sphere, as expected. One can now proceed to solve these equations order by order in $\varepsilon$. For instance, one finds
\begin{align}
\gamma^{(1)}& =\frac{2\ 2^{3/4} \sqrt{7}}{3 \pi^{1/4}}\,,\quad \gamma^{(2)}=-\frac{3686}{189} \sqrt{\frac{2}{\pi }}\,,\quad \gamma^{(3)}=\frac{5022016\ 2^{1/4} }{11907 \sqrt{7} \pi ^{3/4}}\,, \nonumber
\\
q_1^{(1)}(\theta) & =-\frac{2\ 2^{1/4}}{\sqrt{7} \pi ^{3/4}}[2 \cos (4 \theta )+2 \cos(2 \theta )-1]\,,\nonumber
\\
q_1^{(2)}(\theta) & =-\frac{8}{147 \pi } [25 \cos (2 \theta )+25 \cos (4 \theta )-21 \cos (6 \theta )+166]\,, \label{eq:pertell2}
\\
q_4^{(1)}(\theta) & = -\frac{2\ 2^{1/4} \sqrt{7}}{3 \pi ^{3/4}} [2 \cos (2 \theta )+1]\,,\nonumber
\\
q_4^{(2)}(\theta) & = -\frac{2}{189 \pi } [430 \cos (2 \theta )+1727]\,.\nonumber
\end{align}
Note that the fact that $\gamma^{(1)}>0$ indicates that at least for sufficiently small $Q$, the mode that used to go negative for $AdS_2\times S^3$, becomes positive! We shall see that this remains the case for all values of $Q$ we have managed to probe.

For any other perturbation with $\ell\geq 3$ our perturbative scheme starts with nontrivial $\{\gamma^{(0)},q_1^{(0)},q_4^{(0)}(\theta)\}$, since $\gamma^{(0)}$ is non-zero at $Q=0$ for any other value of $\ell\geq3$ (see Fig.~\ref{fig:gamma_negative}). For instance, for the $\ell=3$ mode we find
\begin{equation}
\gamma = \frac{1}{2}-\frac{5 \sqrt{7} }{4 (2 \pi)^{1/4}}\varepsilon+\frac{2987}{672 \sqrt{2 \pi }} \varepsilon ^2-\frac{317057}{6272 \sqrt{7} (2 \pi )^{3/4}} \varepsilon ^3+\mathcal{O}(\varepsilon^4)\,.
\label{eq:pertell3}
\end{equation}

\begin{figure}[th]
\centering \includegraphics[width=0.9\textwidth]{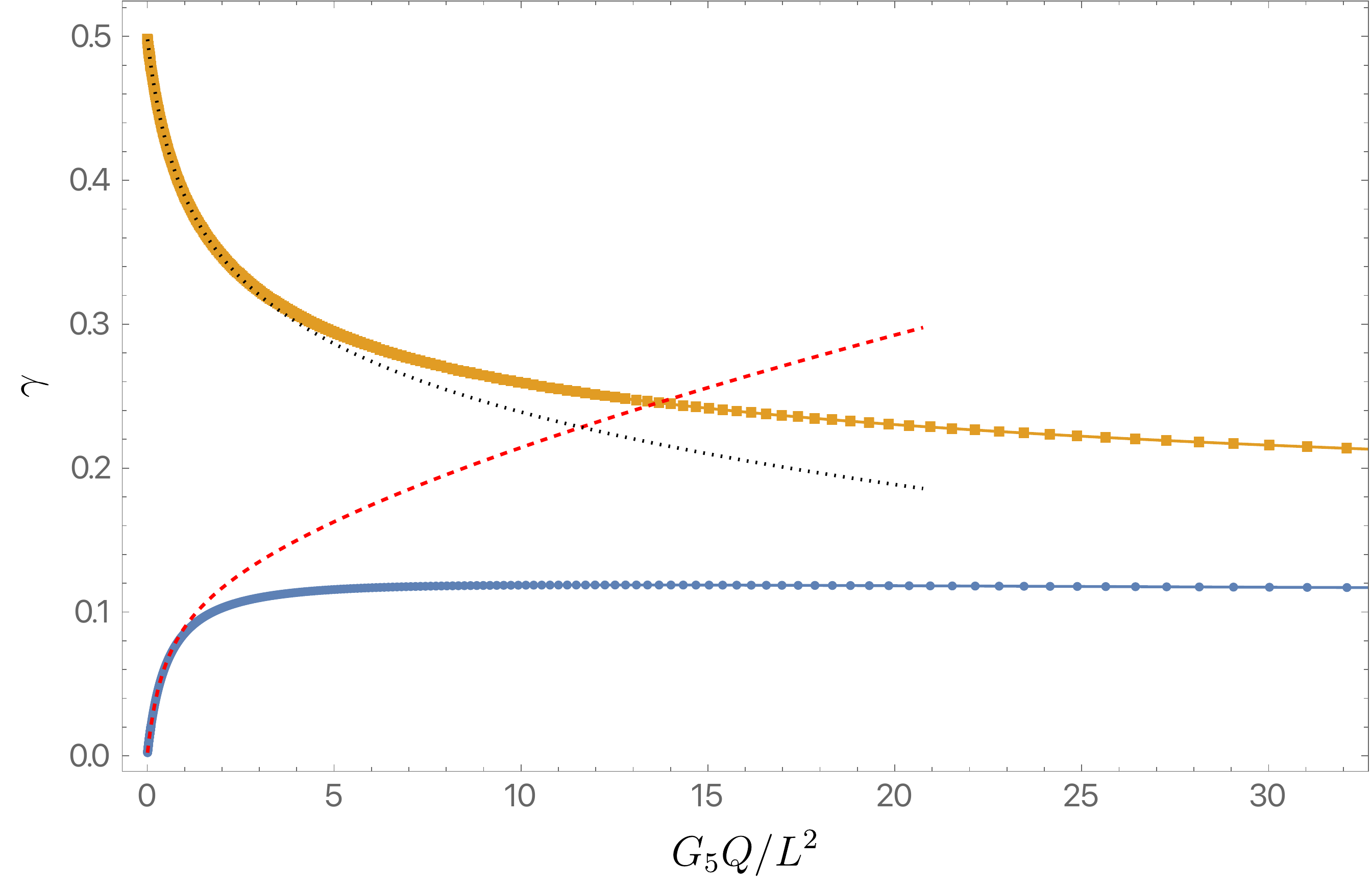}
\caption{\label{fig:per} The two lowest lying modes of the quadratic St\"urm-Liouville equations  Eq.~(\ref{eq:deltaq1}) and Eq.~(\ref{eq:deltaq4}) governing perturbations of the new near horizon geometries. The red dashed line shows the perturbative result displayed in Eq.~(\ref{eq:pertell2}), whereas the black dotted line shows $\gamma$ given in Eq.~(\ref{eq:pertell3}). The blue disks have $\ell=2$ and the orange squares have $\ell=3$. In the language used in section \ref{sec:unstable}, the modes shown both connect to $\gamma_-(\ell,0)$.}
\end{figure}

We now proceed using our exact numerical solutions for the background and by solving Eq.~(\ref{eq:deltaq1}) and Eq.~(\ref{eq:deltaq4}) numerically. We again use the numerical methods detailed in \cite{Dias:2015nua} to do this calculation. The results are displayed in Fig.~\ref{fig:per} where we track the two lowest lying modes. These are associated with $\ell=2$ and $\ell=3$ perturbations, respectively. Recall that for each $\ell\geq2$ perturbation of $AdS_2\times S^3$, there are two physical scaling exponents $\gamma$, which we labelled $\gamma_{\pm}(\ell,y_+)$. The  perturbations we study naturally connect to $\gamma_-(\ell,0)$ for $\ell = 2,3$.  Recall that for each value of $\ell$, horizon boundary conditions allow us to discard two negative values of $\gamma$ (which connect to the values of $\gamma$ that we discard when we plot Fig.~\ref{fig:gamma_negative}). Since the lowest scaling exponent in Fig.~\ref{fig:per} remains positive, this  conclusively shows that the novel IR geometry is RG stable at the linear level to $SO(3)$-symmetric deformations. Furthermore, when $Q$ is small enough, our perturbative results in Eq.~(\ref{eq:pertell2}) (dashed red line) and Eq.~(\ref{eq:pertell3}) (dotted black line) match well our exact numerical results given by the blue disks and orange squares for $\ell=2,3$, respectively.

We tracked the lowest lying mode all the way to $G_5 Q/L^2 \sim 3\times 10^4$ and it remains positive, saturating at around $\gamma \approx 0.1013$. 

\subsection{RG instability of the new IR geometries with respect to $SO(3)$ breaking deformations}
Having investigated the stability properties of the new IR geometries with respect to $SO(3)$-preserving deformations, we now ask whether the new geometries are stable with respect to deformations that break $SO(3)$. Unfortunately, this is \emph{not} the case, as we  show below.

We start by presenting an Ans\"atze for the metric and gauge field perturbations. These are necessarily more involved than the $SO(3)$ symmetric case. Since we want to break the symmetries of the round $S^2$, we expand all perturbations in terms of standard spherical harmonics  $Y_{k\,m}(\chi,\phi)$ on $S^2$, where $\chi\in[0,\pi]$ and $\phi\sim \phi+2\pi$ are the standard latitude and longitude angles on the round $S^2$, respectively. Spherical harmonics on the $S^2$ obey
\begin{equation}
\Box_{\Omega_2}Y_{k\,m}+k(k+1)Y_{k\,m}=0\,,
\end{equation}
with $k=0,1,2,\ldots$ and $|m|\leq k$ being the standard quantum numbers of the spherical harmonics and $\Box_{\Omega_2}$ the standard Laplacian on the round two-sphere. The sector with $k=0$ was studied in the previous section, and the sector with $k=1$ has to be treated separately. For this reason, we take $k \geq2$ from here onward.

There are \emph{two} types of gravito-electromagnetic perturbations we can build from scalar harmonics. This is because vector harmonics on the $S^2$ can be built from Hodge duals of gradients of scalar harmonics (up to harmonic vectors). Perturbations built from scalar harmonics are often called scalar derived gravito-electromagnetic perturbations, while perturbations built from vector harmonics are often coined vector derived gravito-electromagnetic perturbations. The former sector is the one of interest to us, since one can easily show that vector derived gravito-electromagnetic perturbations are RG stable.

Let $\Dbar$ be the metric preserving connection on the round two-sphere, so that $\Dbar_\alpha \Dbar^\alpha = \Box_{\Omega_2}$, with lower case Greek indices running on $S^2$, \emph{i.e.} $\alpha=\{\chi,\phi\}$. We then introduce
\begin{equation}
S^{k\,m}_{\alpha\beta}\equiv \Dbar_{\alpha}\Dbar_{\beta}Y_{k\,m}+\frac{k(k+1)}{2}\,{g}_{\alpha\beta}\,Y_{k\,m}\,,
\end{equation}
where ${g}_{\alpha\beta}$ being the metric on the round two-sphere. By construction, $S_{\alpha\beta}$ is traceless. We then write the following Ans\"atze for the metric and gauge field perturbations
\begin{subequations}
\begin{multline}
\delta \mathrm{d}s^2\equiv h_{ab}\mathrm{d}x^a\mathrm{d}x^b=L^2\rho^\gamma\,\Bigg\{B(\theta)\,\frac{\hat{q}_1(\theta)}{\gamma}\,Y_{k\,m}\left(-A_0^2\,\rho^2\,\frac{\mathrm{d}t^2}{L^2}+\frac{\mathrm{d}\rho^2}{\rho^2}\right)
\\
+Y_+^2\Big[H(\theta)^2\, \hat{q}_2(\theta)\,Y_{k\,m}\,\mathrm{d}\theta^2+\frac{\sin^2\theta}{H(\theta)}\hat{q}_3(\theta)\,Y_{k\,m}\,\mathrm{d}\Omega_2^2
\\
+2\,\frac{\hat{q}_5(\theta)}{\gamma\,k}\,\,\mathrm{d}\theta\,(\Dbar_{\alpha}Y_{k\,m})\mathrm{d}x^\alpha+\hat{q}_6(\theta)\,S^{k\,m}_{\alpha\beta}\mathrm{d}x^\alpha\,\mathrm{d}x^\beta\Big]\Bigg\}\,,
\end{multline}
and
\begin{equation}
\delta A \equiv a_a \mathrm{d}x^a = -\rho_{\rm IR}\,A_0\,\rho^{1+\gamma}\,\hat{q}_4(\theta)\,Y_{k\,m}\,\mathrm{d}t\,.
\end{equation}
\end{subequations}
There are total of six functions of $\theta$ to solve for, namely $\{\hat{q}_1,\ldots,\hat{q}_6\}$. After some considerable algebra, one can express $\hat{q}_2$, $\hat{q}_3$ and $\hat{q}_6$ as a function of the remaining unknown functions and their first derivatives with respect to $\theta$. We are thus left with three second order ordinary differential equations in $\theta$ for $\{\hat{q}_1,\hat{q}_4,\hat{q}_5\}$. Regularity at the poles demands
\begin{equation}
\hat{q}_1(\theta)\approx \sin^k \theta\,\hat{C}_1\,,\quad \hat{q}_4(\theta)\approx \sin^k\theta\,\hat{C}_2 \quad\text{and}\quad \hat{q}_5(\theta)\approx \sin^{k-1} \theta\,\hat{C}_3\,,
\end{equation}
where $\hat{C}_1$, $\hat{C}_2$ and $\hat{C}_3$ are constants. In order to impose these, we change to a new set of variables
\begin{equation}
\hat{q}_1(\theta)=\sin^k \theta\,\hat{Q}_1(\theta)\,,\quad \hat{q}_4(\theta)=\sin^k \theta\,\hat{Q}_2(\theta)\quad\text{and}\quad \hat{q}_5(\theta)=\sin^{k-1} \theta\,\hat{Q}_3(\theta)\,
\end{equation}
with regularity at the poles now simply demanding
\begin{equation}
\hat{Q}_1^\prime(0)=\hat{Q}_2^\prime(0)=\hat{Q}_3^\prime(0)=\hat{Q}_1^\prime(\pi)=\hat{Q}_2^\prime(\pi)=\hat{Q}_3^\prime(\pi)=0\,.
\end{equation}

It is possible to cast, with the above boundary conditions, the second order differential equations for $\hat{Q}_1$, $\hat{Q}_2$ and $\hat{Q}_3$ as a Sturm-Liouville problem, where the combination $\gamma(\gamma+1)$ appears as the eigenvalue. This is the system we solve numerically using the numerical methods detailed in \cite{Dias:2015nua}.

\begin{figure}[th]
\centering \includegraphics[width=0.9\textwidth]{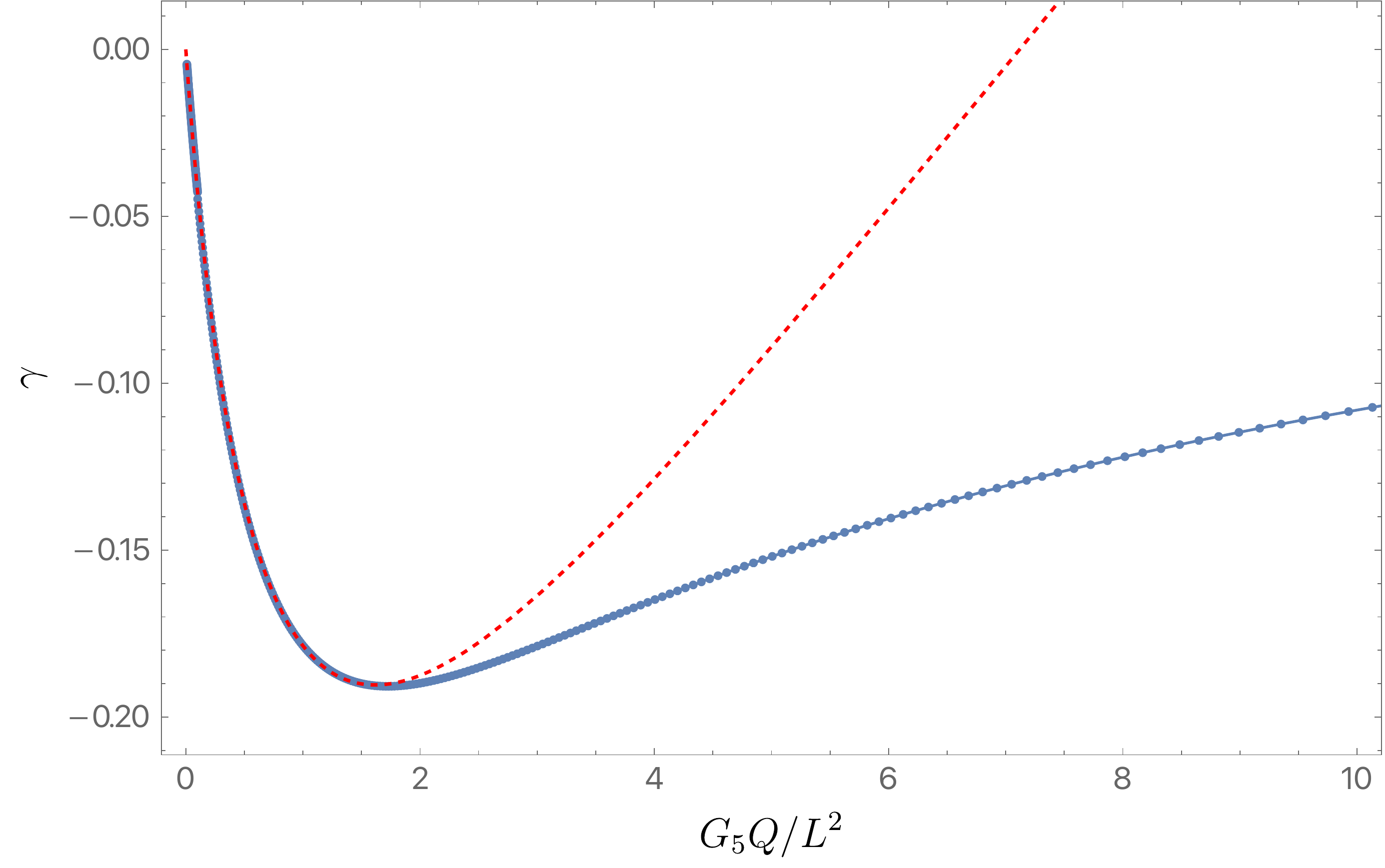}
\caption{\label{fig:so3_violating} The lowest lying scaling exponent $\gamma$ as a function of $ Q$ for $SO(3)$ breaking perturbations. The dashed red line is the perturbative result (\ref{eq:so3_viol_per})  whereas the blue disks label the exact numerical results. The agreement between the two at small charges is reassuring for both methods.}
\end{figure}

We start by using the perturbative scheme of section \ref{sec:ell2}, which allows us to predict $\gamma$ for small enough charge $Q$. Indeed, for $k=2$ we find that 
\begin{equation}
\gamma=-\frac{4\ 2^{3/4} \sqrt{7} }{3 \pi^{1/4}}\varepsilon+\frac{496}{27} \sqrt{\frac{2}{\pi }} \varepsilon^2+\frac{6259984 \ 2^{1/4}}{11907 \sqrt{7} \pi
   ^{3/4}} \varepsilon^3+\frac{52933540168}{26254935 \pi } \varepsilon ^4+\mathcal{O}(\varepsilon^5)\,.
\label{eq:so3_viol_per}
\end{equation}
The fact that the first term in the $\varepsilon$ expansion of $\gamma$ is negative (note that we must take $\varepsilon>0$ in order for our perturbative scheme to make sense) is a signal that our new geometries are RG unstable with respect to $SO(3)$ breaking perturbations. The question remains as to whether for larger values of the charge (or alternatively, larger values of $\varepsilon$) $\gamma$ will become positive. In order to address this question we solve the problem numerically, and report our findings in Fig.~\ref{fig:so3_violating}. As a dashed red line we plot our perturbative result (\ref{eq:so3_viol_per}), while the exact numerical results are shown as blue disks. The agreement between the two at small charges is reassuring. The fact that $\gamma$ remains \emph{negative} for all values of the charge is the main result of this section and shows that our $SO(3)$ symmetric zero-temperature geometries are RG unstable to perturbations that break $SO(3)$. The endpoint of this $SO(3)$ breaking instability remains unknown and is under current investigation.

\section{Approaching the new IR at low temperature}
Having studied the RG stability linearly, we now proceed to a fully nonlinear treatment. Rather than focus on the near horizon geometry as we have done so far, in this section we construct asymptotically AdS black holes with a finite $SO(3)$ (but not $SO(4)$) invariant  deformation on the boundary.  Our metric and gauge field Ansatz closely follows the ones used in \cite{Costa:2015gol,Costa:2017tug}, but adapted to $D=5$.

To make progress at the nonlinear level, we use the DeTurck method. This method was first proposed in \cite{Headrick:2009pv,Adam:2011dn} and reviewed in \cite{Wiseman:2011by,Dias:2015nua}. It is a very hard problem (but a very interesting one) to work directly at zero temperature. We bypass this problem by working at a finite temperature $T$ and lower $T$ as much as possible.  We will provide numerical evidence in favour of flows that start at the boundary with a particular $SO(3)$-invariant boundary deformation, and approach in the deep IR the solutions described in section \ref{sec:exact}. 
Nearby deformations behave similarly, so an open set of $SO(3)$-invariant boundary data exists with a near horizon geometry that approaches our novel IR solutions as $T\to 0$.

We will insist on having a conformal boundary metric that is conformal to the Einstein static universe. As such, we take
\begin{equation}
\mathrm{d}s^2_{\partial}= -\mathrm{d}t^2+L^2(\mathrm{d}\theta^2+\sin^2\theta\,\mathrm{d}\Omega_2^2)\,.
\end{equation}
For the gauge field, we will take a chemical potential that depends on $\theta\in[0,\pi]$ only, \emph{i.e.}
\begin{equation}
A_\partial=\mu(\theta)\mathrm{d}t\,.
\end{equation}
We will also insist that our bulk metric preserves spherical symmetry (with respect to the $S^2$) and be static with respect to $\partial/\partial t$, so that $\partial/\partial t$ is hypersurface orthogonal in the bulk. 

\subsection{Numerical method}
With the above symmetries, the most general Ansatz that we can write is given by
\begin{subequations}
\begin{multline}
\mathrm{d}s^2=\frac{L^2}{(1-y^2)^2}\Bigg\{-G(y) y^2\frac{A_1(x,y)}{L^2}\mathrm{d}t^2+\frac{4\,y_+^2}{G(y)}A_2(x,y)\left[\mathrm{d}y+A_3(x,y)\mathrm{d}y\right]^2
\\
+y_+^2\left[\frac{4\ A_4(x,y)\mathrm{d}x^2}{2-x^2}+A_5(x,y)(1-x^2)^2\,\mathrm{d}\Omega_2^2\right]\Bigg\}\,,
\label{eq:ansat}
\end{multline}
and
\begin{equation}
A=A_6(x,y)\ y^2 \mathrm{d}t\,,
\end{equation}
with
\begin{equation}
G(y)\equiv \left(2-y^2\right) \left[\left(1-y^2\right)^2+\left(2-2 y^2+y^4\right) y_+^2-\frac{4}{3} \left(1-y^2\right)^4 \tilde{\mu} ^2\right]\,.
\end{equation}
\label{eq:ansatznonlinearly}
\end{subequations}
For $\mu(\theta)$ we take
\begin{subequations}
\begin{equation}
\mu(\theta)=\bar{\mu}+\sum_{\ell=1}^{+\infty} \mu_{\ell} \ Y_{\ell}(\theta)\,,
\label{eq:boundarychemical}
\end{equation}
with $Y_{\ell}(\theta)$ harmonics on the three-sphere normalised so that $Y_{\ell}(0)=1$, and explicitly given by
\begin{equation}
Y_{\ell}(\theta) = \frac{1}{\ell+1}\frac{\sin [(\ell +1) \theta ]}{ \sin \theta}\,.
\end{equation}
\end{subequations}%
The relation between $\theta$ and $x$ is given by
\be\label{eq:theta}
\cos \theta = x\sqrt{2-x^2}
\ee{}

To gain some intuition for the above form of the metric, let us imagine for a moment that all $\mu_{\ell}=0$ for $\ell\geq1$. If we take
\begin{equation}
A_1=A_2=A_4=A_5=1\,,\quad A_6=(2-y^2)\bar{\mu}\,,\quad A_3=0\quad  \text{and} \quad \tilde{\mu}=\bar{\mu}
\label{eq:reference}
\end{equation}
then Eqs.~(\ref{eq:ansatznonlinearly}) describe a five-dimensional Reissner-Nordstr\"om black hole with radius $r_+\equiv y_+ L$ and chemical potential $\bar{\mu}$. To see this we do a simple change of coordinates
\begin{equation}
 r = \frac{r_+}{1-y^2}\,,
\label{eq:RNo}
\end{equation}
which brings Eq.~(\ref{eq:ansatznonlinearly}) to the more familiar form
\begin{subequations}
\begin{equation}
\mathrm{d}s^2_{\rm RN} = -f(r)\mathrm{d}t^2+\frac{\mathrm{d}r^2}{f(r)}+r^2(\mathrm{d}\theta^2+\sin^2\theta \mathrm{d}\Omega_2^2)\quad \text{and}\quad A = \bar{\mu}\left(1-\frac{r_+^2}{r^2}\right)\mathrm{d}t\,,
\end{equation}
with
\begin{equation}
f(r)=\frac{r^2}{L^2}+1-\frac{r_+^2}{r^2}\left(1+\frac{r_+^2}{L^2}+\frac{4}{3}\bar{\mu}^2\right)+\frac{4}{3}\bar{\mu}^2\frac{r_+^4}{r^4}\,.
\end{equation}
\end{subequations}

When using the DeTurck method, one has to choose a reference metric $\bar{g}$ that will ultimately fix the gauge. For the reference metric $\bar{g}$ we take $A_i$, with $i=\{1,\ldots,5\}$ as in Eq.~(\ref{eq:reference}). One then solves the Einstein-DeTurck equation, which are given by
\begin{equation}
R_{ab}+\frac{4}{L^2}g_{ab}-\nabla_{(a}\xi_{b)}=2\left(F_{a}^{\phantom{a}c}F_{bc}-\frac{g_{ab}}{6}F_{cd}F^{cd}\right)\,,
\label{eq:turck}
\end{equation}
with $\xi^a = \left[\Gamma^a_{bc}(g)-\Gamma^a_{bc}(\bar{g})\right]g^{cd}$, where $\Gamma(\mathfrak{g})^a_{cd}$ are the components of the standard Christoffel symbol associated with a metric $\mathfrak{g}$. Stationary solutions of the Einstein-DeTurck equation in vacuum can be shown to coincide with genuine solutions of the Einstein equation \cite{Figueras:2011va,Figueras:2016nmo}, that is to say, on solutions of the Einstein DeTurck equation, $\xi=0$.

However, in the presence of matter, such a proof does not exist, and \emph{a priori} Ricci solitons, \emph{i.e.} solutions with $\xi\neq0$, cannot be ruled out. Instead, it has been first noted in \cite{Horowitz:2012ky} that Eq.~(\ref{eq:turck}) represents an Elliptic system of equations for stationary metrics of the form Eq.~(\ref{eq:ansat}), even in the presence of matter,  once appropriate boundary conditions are imposed.  This allows us to show that the solutions we construct are not Ricci solitons, as we explain below.

At the conformal boundary, located at $y=1$, we demand that the metric approaches the reference metric, so that
\begin{subequations}
\begin{equation}
A_1(x,1)=A_2(x,1)=A_4(x,1)=A_5(x,1)=1\quad \text{and}\quad A_3(x,1)=0\,.
\end{equation}
For the Maxwell field we take instead 
\begin{equation}
A_6(x,1)=\mu(\theta)
\end{equation}
\end{subequations}%
with $\mu(\theta)$ given in Eq.~(\ref{eq:boundarychemical}) and $x$ related to $\theta$ as in Eq.~(\ref{eq:theta}). 

At the bifurcating Killing three-sphere, located at $y=0$, we demand
\begin{align}
&A_1(x,0)=A_2(x,0)\,,\quad A_3(x,0)=0\,,\nonumber
\\
&\left.\frac{\partial A_2}{\partial y}\right|_{y=0}=\left.\frac{\partial A_4}{\partial y}\right|_{y=0}=\left.\frac{\partial A_5}{\partial y}\right|_{y=0}=\left.\frac{\partial A_6}{\partial y}\right|_{y=0}=0\,.
\end{align}
The first condition fixes the black hole temperature to be 
\begin{equation}
T = \frac{3+6 y_+^2-4 \tilde{\mu }^2}{6 \pi  L y_+}\,.
\end{equation}
The equation above reveals the physical meaning of $y_+$ and $\tilde{\mu}$: these are parameters that we can use to dial the temperature. Note that these are redundant in the sense that we have two parameters to set a single number, the temperature. Additionally, note that the reference metric sets the gauge, and as such $y_+$ and $\tilde{\mu}$ also control our gauge choice.

We are still left with detailing the boundary conditions at the axis of symmetry $\theta=0,\pi$ or equivalently $x=\pm1$. Here we want the metric to be smooth, and that is equivalent to demanding
\begin{align}
&A_4(\pm 1,y)=A_5(\pm 1,y)\,,\quad A_3(\pm 1,y)=0\,,\nonumber
\\
&\left.\frac{\partial A_1}{\partial x}\right|_{x=\pm 1}=\left.\frac{\partial A_2}{\partial x}\right|_{x=\pm 1}=\left.\frac{\partial A_5}{\partial x}\right|_{x=\pm 1}=\left.\frac{\partial A_6}{\partial x}\right|_{x=\pm 1}=0\,.
\end{align}

We thus see that our problem naturally lives on a square integration domain $(x,y)\in[0,1]^2$. It is a simple exercise  to show that the Einstein-DeTurck equation Eq.~(\ref{eq:turck}), together with the boundary conditions above and restricted to the line element Eq.~(\ref{eq:ansat}) does give rise to a well posed Elliptic problem, which we can now solve using the numerical methods of \cite{Dias:2015nua}. Additionally, since Elliptic equations admit locally unique solutions, one can differentiate between a Ricci soliton and a \emph{bona fide} solution of the Einstein equation by monitoring $\xi^a$. Furthermore, since the $(x,y)$ base space is manifestly positive definite (as it should in an Elliptic problem) we can equally well monitor instead $\xi_a \xi^a$. To solve this problem we used a pseudo-spectral collocation methods.

\begin{figure}[th]
\centering \includegraphics[width=0.6\textwidth]{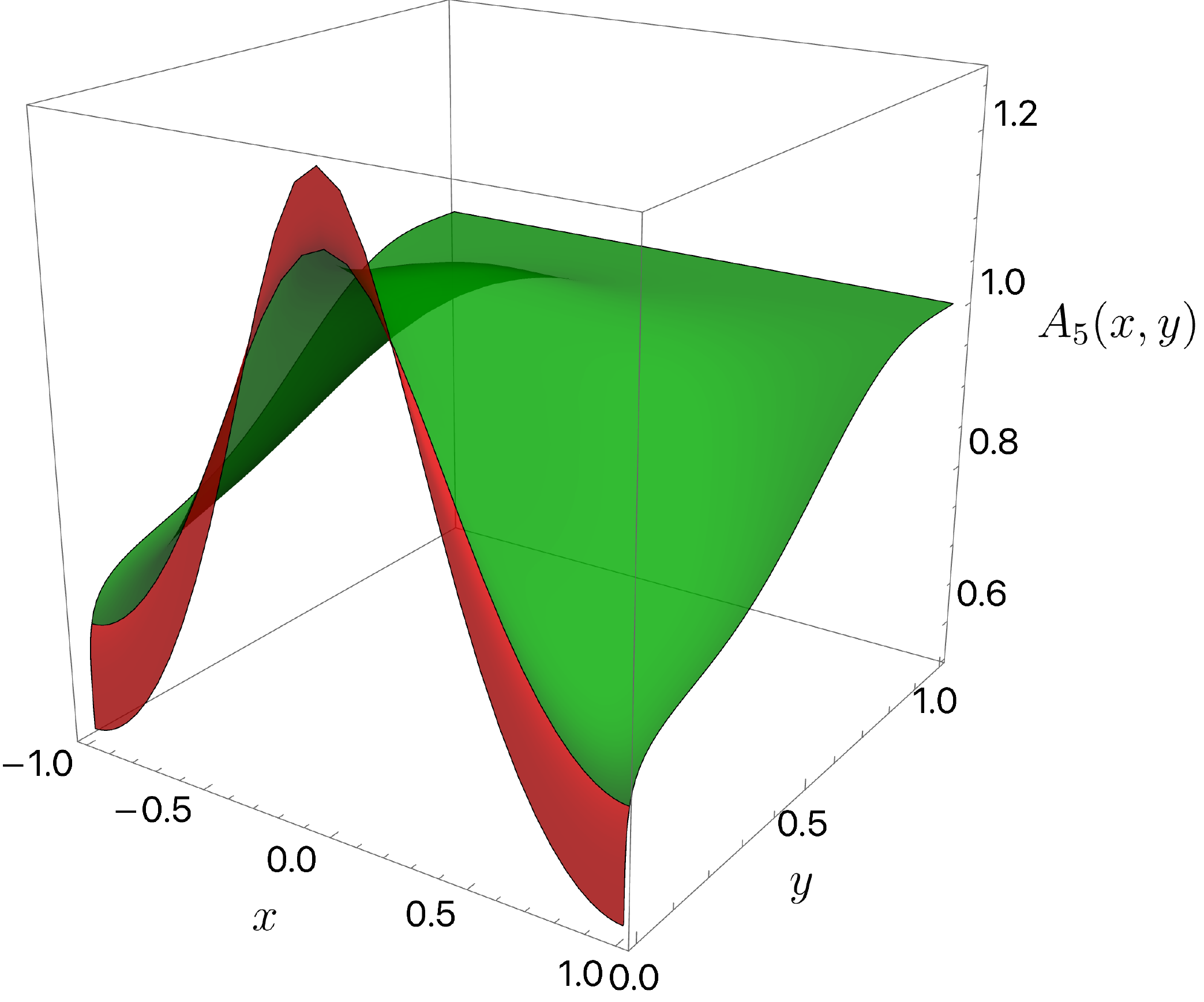}
\caption{\label{fig:domain} A plot showing $A_5(x,y)$ as a function of $(x,y)$ with $4 \pi L T =10^{-8}$, $\bar{\mu}=1.26244$, $\mu_2=-0.5$ and $\mu_{\ell}=0$, for $\ell \neq 2$. In this case $y_c=10^{-2}$, and the patch near the horizon is represented in red and the one away from the horizon in green.}
\end{figure}

As we shall see, we will be interested in reaching remarkably low temperatures. At really low temperatures, enormous gradients develop near the bifurcating Killing three-sphere. To deal with these we discretize our equations on \emph{two} Chebyshev-Gauss-Lobatto grids, which connect at an interface $0<y_c<1$ (see \cite{Dias:2015nua} on more details on how to deal with patching procedures). In our numerical simulations we take $y_c$ to be small, so that the new grid covers most of the gradients at the horizon. Typically, $y_c\sim10^{-2}$ and we use not less than $50\times 50$ collocation points on each subdomain. In Fig.~\ref{fig:domain} we show a plot of $A_5$ (which is the radius of the round $S^2$, and is gauge invariant) as a function of $(x,y)$. This particular image was generated with $4 \pi L T =10^{-8}$, $\bar{\mu}=1.26244$, $\mu_2=-0.5$ and $\mu_{\ell}=0$, for $\ell \neq 2$ and with $y_c=10^{-2}$. In the plot, large gradients near the horizon are easily identifiable and justify the use of patching. We tried using a single domain, and it proved almost impossible to reach such low temperatures, partially because such a large number of points was needed that even with extended precision it was hard to run a Newton-style method to solve the resulting discretized equations.

\subsection{Results}

The aim of this section is show that the near horizon geometries that we constructed in section \ref{sec:exact} do appear as the IR of a class of boundary chemical potentials of the form given in Eq.~(\ref{eq:boundarychemical}). We will first focus our attention on profiles with $\mu_{\ell}=0$ for $\ell\neq2$, and say a few words about the general case near the end of this section.

To understand the approach to the near horizon geometries of \ref{sec:exact} we need a reference, with the most standard being the entropy. However, as we can see in Fig.~\ref{fig:large}, we need to get to very large charges in order to see large deviations with respect to $AdS_2\times S^3$. So, instead of comparing directly with the entropy $S$, we are going to compare with $\Delta S$. Recall that $\Delta S$ measures the difference in entropy between  a given solution and that of an extreme Reissner-Nordstr\"om with the same total charge.

\begin{figure}[th]
\centering \includegraphics[width=0.9\textwidth]{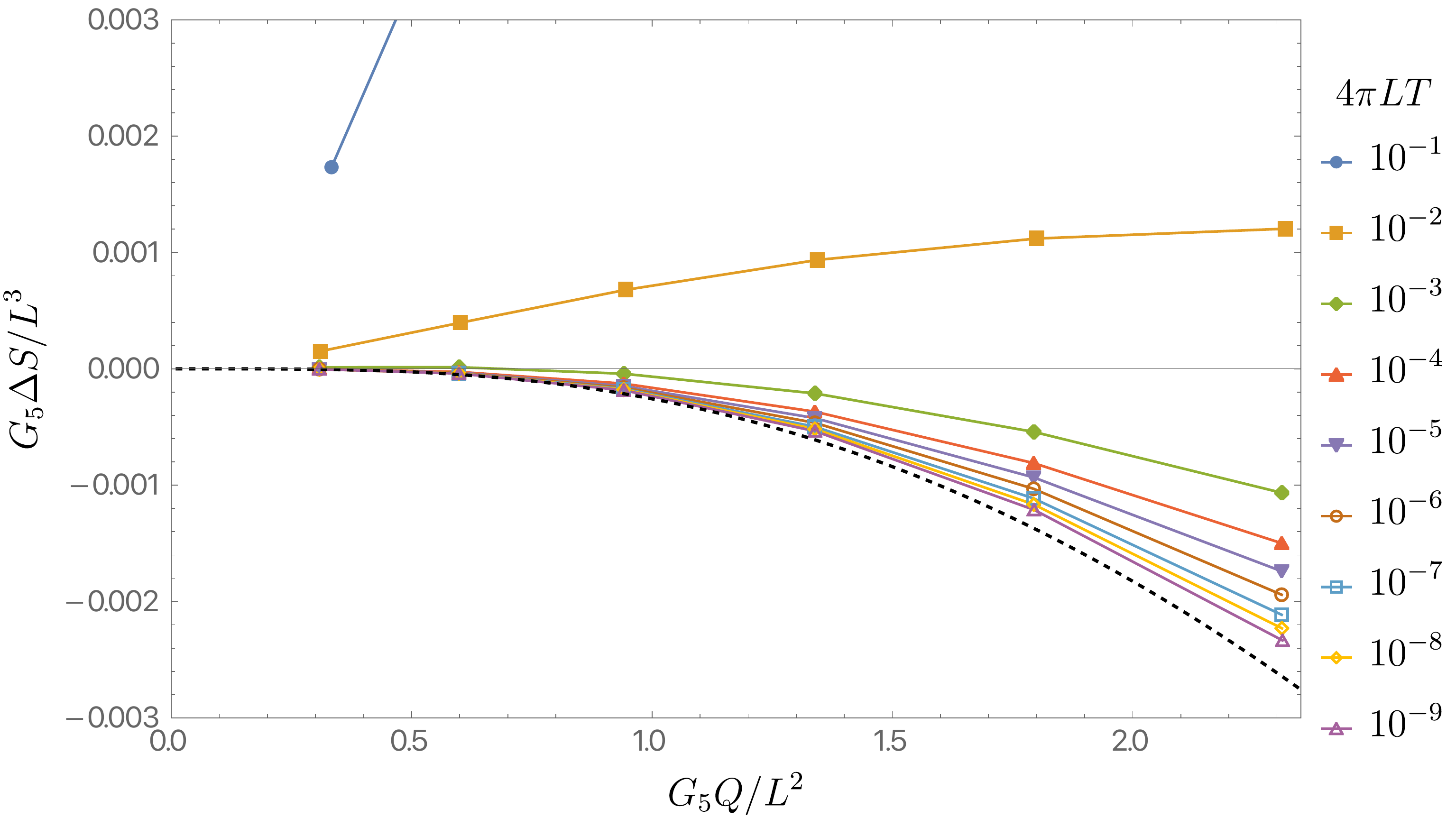}
\caption{\label{fig:figuremerit} $\Delta S$ as a function of $Q$ computed for several fixed temperatures shown on the figure caption on the right. The black dashed line shows the entropy appearing in Fig.~\ref{fig:large} for the $T=0$ near horizon geometry (for charges up to  $G_5 Q/L^2\sim 2$).}
\end{figure}

In Fig.~\ref{fig:figuremerit} we plot $\Delta S$ as a function of $Q$ for several fixed temperatures, indicated on the figure. To generate each curve, we fix the temperature and vary $\bar{\mu}$ while keeping $\mu_2=-0.5$. The black dashed line corresponds to $\Delta S$ for the $T=0$ near horizon geometry as shown in  Fig.~\ref{fig:large} (for charges up to  $G_5 Q/L^2\sim 2$). It is clear that, as we lower $T$, we approach the $\Delta S$ shown in Fig.~\ref{fig:large}.

\begin{figure}[th]
\centering \includegraphics[width=0.6\textwidth]{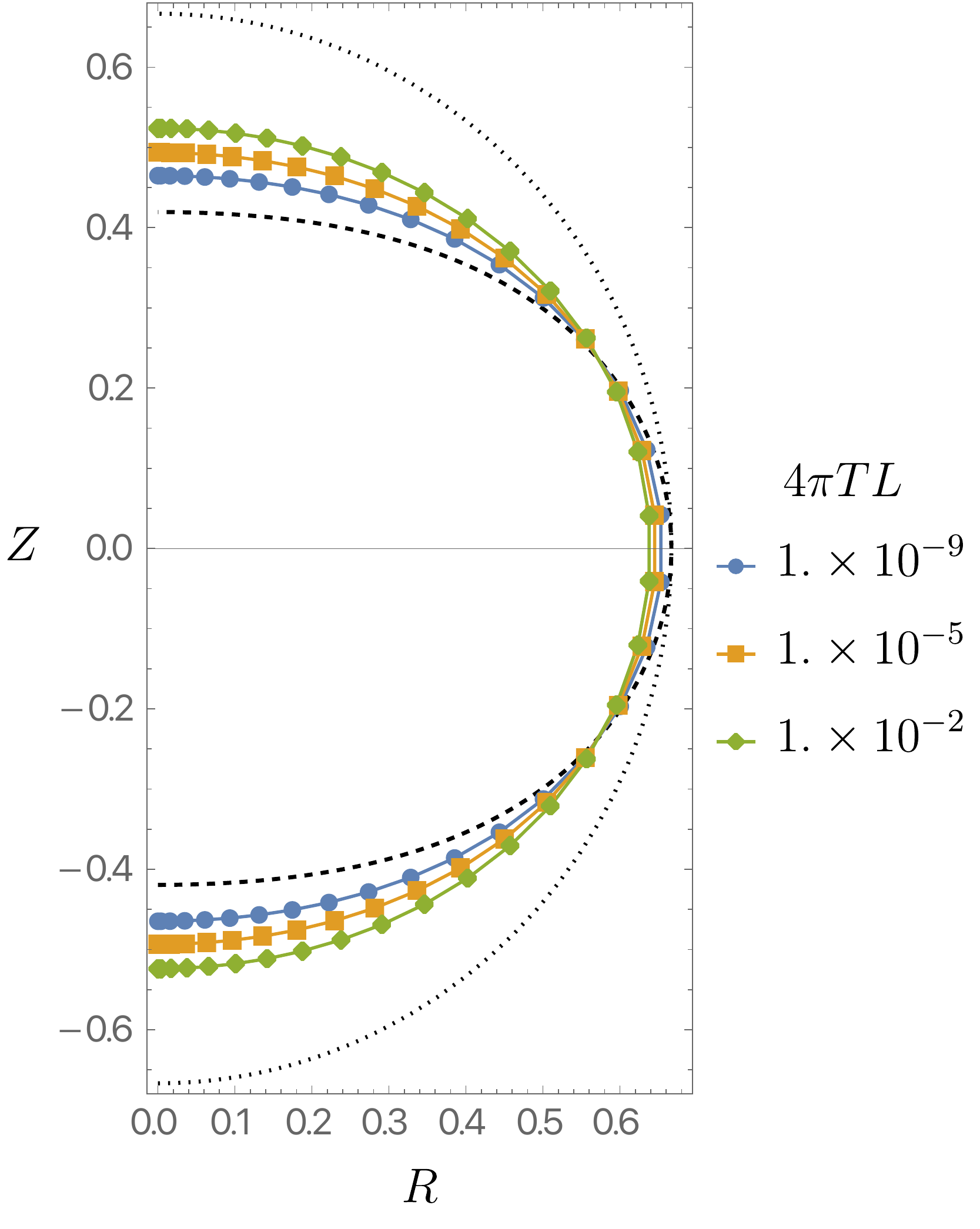}
\caption{\label{fig:local} Isometric embedding of the finite temperature horizon geometry into $\mathbb{R}^4$. The black dotted line shows what a perfect sphere would look like and the dashed line represents our novel pancaked IR geometry discussed in section \ref{sec:exact}. The several symbols represent the embeddings of our finite temperature solutions, whose temperatures are labeled on the right. The solution with $4\pi T L = 10^{-9}$ has $G_5 Q/L^2\approx 1.33896$, whereas the exact zero temperature solution has $G_5 Q/L^2\approx 1.33653$ (corresponding to an agreement of about $0.18\%$).}
\end{figure}

In addition to monitoring $\Delta S$, we also looked into some detail on the local horizon geometry. In particular, we embedded spatial cross sections of our finite temperature horizons and compared those with spatial cross sections of our zero-temperature horizons of section \ref{sec:exact} (see for instance Fig.~\ref{fig:embedding}). The results of this comparison are presented in Fig.~\ref{fig:local}.

Finally, we make a few comments about more general profiles for $\mu(\theta)$. So far, we have discussed the result of just turning on $\mu_{2}$, an $\ell =2$ contribution to $\mu(\theta)$.  We have checked that the overall picture does not change if we turn on other harmonics so long as $|\mu_{\ell}|\ll |\mu_{2}|$ for $\ell\neq2$. This is not surprising since even when we turn on a single harmonic at the boundary, all other harmonics are nonlinearly generated in the bulk.
For all the plots shown we have chosen $\mu_2=-0.5$, but we have tried other values of $\mu_2<0$ and found similar results. So we indeed have an open set of $SO(3)$-invariant chemical potentials that flow to our new IR geometries. For $\mu_2>0$ we do not have enough numerical evidence to provide a clear picture. At the moment it appears that the near horizon geometries of section \ref{sec:exact} are not approached as we lower $T$ with this sign of the deformation.

\section{General nonrotationally invariant near horizon geometries}

We have also found novel near horizon geometries where the symmetries of the three-sphere are completely broken.   It was shown in \cite{Kunduri:2007vf} that (smooth) static near horizon geometries must be a warped product of $AdS_2$ and a compact ($D-2)$-manifold, so we keep the  $AdS_2$ symmetries.   We have found no such geometries for arbitrarily small charges, \emph{i.e.} near $Q=0$. However, from Eq.~(\ref{eq:yc}), it is clear that novel solutions exist for $\ell\geq3$ that are perturbatively close $AdS_2\times S^3$. These branch from $AdS_2\times S^3$ precisely when $y_+=y_+^c(\ell)$, with $y_+^c(\ell)$ given in Eq.~(\ref{eq:yc}).

One might then ask how many solutions do we expect to branch from $AdS_2\times S^3$ at a given $y_+=y_+^c(\ell)$. We can answer this by determining the number of parameters in the most general harmonic of degree $\ell$. The most general harmonic of degree $\ell$ will depend on $(\ell+1)^2$ parameters (this corresponds to the degeneracy of spherical harmonics on the three-sphere). However, some of these harmonics can be related by acting with $SO(4)$ \emph{i.e.} by rotations of the background $S^3$. Since $SO(4)$ has dimension $6$, we can eliminate $6$ parameters. This leaves us with $(\ell+1)^2-6$ parameters that cannot be eliminated via background rotations,  and thus with $(\ell+1)^2-6$ new near horizon geometries branching out of each $y_+=y_+^c(\ell)$. (Note that in this section we are taking $\ell\geq3$.)

For even values of $\ell$, the story is more intricate, because it turns out that a perturbation with an overall amplitude $A$ cannot be mapped into a perturbation with an amplitude $-A$ via a background symmetry. This means that for even values of $\ell$ we  expect $2[(\ell+1)^2-6]$ new near horizon solutions branching out of each $y_+=y_+^c(\ell)$. For odd values of $\ell$, we can use a reflection around the equator to map solutions with an amplitude $A$ into solutions with an amplitude $-A$, and as such we do expect $(\ell+1)^2-6$ solutions near the onset.

We now present the method we used to construct solutions with $\ell=4$ that break all rotational symmetries. From the paragraphs above, we would expect $38$ distinct solutions. However, for the purposes of this section, we restrict ourselves to showing that a solution exists that breaks all rotational symmetries. An exhaustive study that determines all of the $38$ solutions is beyond the scope of this work.

Since the spectrum is degenerate, \emph{i.e.} we have more than one spherical harmonic with the same value of $\ell$, we have to proceed using degenerate perturbation theory. The most general harmonic with $\ell=4$ depends on $25$ distinct parameters. Let us start by writing the $S^3$ in hyperspherical coordinates $\{x=\cos \theta,y=\cos \chi,\phi\}$
\begin{equation}
\mathrm{d}\Omega_3^2=\frac{\mathrm{d}x^2}{1-x^2}+(1-x^2)\left[\frac{\mathrm{d}y^2}{1-y^2}+(1-y^2)\mathrm{d}\phi^2\right]\,.
\end{equation}

In these coordinates, scalar spherical harmonics on the three-sphere can be written as
\begin{equation}
\frac{Y_{\ell\,k\,m}(x,y,\phi)}{N_{\ell\,k\,m}}=\left\{\begin{array}{ll}
(1-x^2)^{\frac{k}{2}}{}_2F_1\left(-\ell+k;\ell+k+2;k+\frac{3}{2};\frac{1-x}{2}\right)\,P_{k}^m(y)\cos(m\,\phi)\,,& m>0
\\
(1-x^2)^{\frac{k}{2}}{}_2F_1\left(-\ell+k;\ell+k+2;k+\frac{3}{2};\frac{1-x}{2}\right)\,P_{k}(y)\,,& m=0
\\
(1-x^2)^{\frac{k}{2}}{}_2F_1\left(-\ell+k;\ell+k+2;k+\frac{3}{2};\frac{1-x}{2}\right)\,P_{k}^m(y)\sin(m\,\phi)\,,& m<0
\end{array}\right.
\end{equation}
where ${}_2F_1(a;b;c;z)$ is a Hypergeometric function, $P_{k}^m(y)$ is an associated Legendre polynomial, $P_k(y)\equiv P_{k}^0(y)$ is a Legendre polynomial and
\begin{equation}
N_{\ell\,k\,m}=\frac{\sqrt{\ell +1} \sqrt{(\ell +k+1)!}}{\sqrt{\pi }\,2^{k+\frac{1}{2}}\, \sqrt{(\ell -k)!}\,\Gamma\left(k+\frac{1}{2}\right) \sqrt{k+\frac{1}{2}}} \sqrt{\frac{(k-m)!}{(k+m)!}}\times
\left\{\begin{array}{l}\frac{1}{\sqrt{2}}\quad \text{if}\quad m=0
\\
1\quad \text{if}\quad m\neq 0\end{array}
\right.
\end{equation}
so that
\begin{equation}
\int_0^{2\pi}\int_{-1}^1\int_{-1}^1\,\sqrt{1-x^2}\,Y_{\tilde{\ell}\,\tilde{k}\,\tilde{m}}(x,y,\phi)\,Y_{\ell\,k\,m}(x,y,\phi)\mathrm{d}\phi\,\mathrm{d}x\,\mathrm{d}y=\delta_{\tilde{\ell}\,\ell}\delta_{\tilde{k}\,k}\delta_{\tilde{m}\,m}\,.
\end{equation}
Additionally, we have $0\leq k\leq\ell$ and $|m|\leq k$, which indeed gives $(\ell+1)^2$ harmonics for a given value of $\ell$, as claimed above.

We now present our metric and gauge field Ans\"atze for these generic configurations 
\begin{subequations}
\begin{multline}
\mathrm{d}s^2=L^2\Bigg\{H_1\,\Bigg(-\frac{\,\rho^2\,\mathrm{d}t^2}{L^2}+\frac{\mathrm{d}\rho^2}{\rho^2}\Bigg)+\frac{H_2}{1-x^2}\left(\mathrm{d}x+H_5\,\mathrm{d}y+H_6\,\mathrm{d}\phi\right)^2
\\
+(1-x^2)\left[\frac{H_3}{1-y^2}\left(\mathrm{d}y+H_7\,\mathrm{d}\phi\right)^2+(1-y^2)H_4\,\mathrm{d}\phi^2\right]\Bigg\}\,,
\label{eq:einstein3D}
\end{multline}
\begin{equation}
A = -\rho_{\rm IR}\,\rho\,\mathrm{d}t\,,
\label{eq:maxwell3D}
\end{equation}
\end{subequations}
where $H_i$ (with $i=1\ldots7$) are functions of $(x,y,\phi)$ to be determined in what follows. The Maxwell equation is automatically satisfied via (\ref{eq:maxwell3D}), but we are still left to solve the Einstein equation. In order to solve the Einstein equation we again use the DeTurck trick and take as our reference metric (\ref{eq:einstein3D}) with $H_5=H_6=H_7=0$, $H_2=H_3=H_4=1$ and $H_1=1$. Note that the Einstein-DeTurck equation depends only on $\rho_{\rm IR}^2$.

{To solve the problem at the nonlinear level we used spectral methods. The idea is to expand the metric as an infinite sum of three types of harmonic blocks: harmonic tensors, harmonic vectors and harmonic scalars. These can be found in great detail \cite{Lindblom:2017maa}. All boundary conditions following from regularity are automatically imposed. It then remains to extract the coefficients in these expansions, which we determine via a Newton-Raphson procedure (more details on the numerical method will appear elsewhere \cite{Santos:2023}).}



We solved the Einstein-DeTurck equation using both perturbative methods and fully non-perturbative numerical solutions with the two methods agreeing well when their regime of validity overlaps

We setup our perturbation theory as follows:
\begin{subequations}
\begin{align}
&H_i(x,y,\phi)=\sum_{I=0}^{+\infty}\varepsilon^I h^{(I)}_i(x,y,\phi)\,,
\\
&C_{\rm IR}\equiv \rho_{\rm IR}^2=\sum_{I=0}^{+\infty}\varepsilon^I C_{\rm IR}^{(I)}\,,
\end{align}
\end{subequations}%
with $h^{(0)}_5=h^{(0)}_6=h^{(0)}_7=0$, $h^{(0)}_2=h^{(0)}_3=h^{(0)}_4=h_0$ and $h^{(0)}_1=a_0$. At zeroth order in $\varepsilon$, we find
\begin{subequations}
\begin{equation}
a_0=\frac{1}{4}\frac{h_0}{1+3h_0}
\end{equation}
and
\begin{equation}
C_{\rm IR}^{(0)}=\frac{3\,h_0}{16}\frac{1+2 h_0}{\left(1+3h_0\right)^2}\,.
\end{equation}
\end{subequations}

At first order in $\varepsilon$, we find
\begin{subequations}
\begin{equation}
h_5^{(1)}=h_6^{(1)}=h_7^{(1)}=0\,,\quad h_2^{(1)}=h_3^{(1)}=h_4^{(1)}=p(x,y,\phi)\quad\text{and}\quad h_1^{(1)}=a(x,y,\phi)\,
\end{equation}
with
\begin{align}
&\hat{\Box} a+8\left(1+h_0\right)a-\frac{32}{3}\left(1+3h_0\right)C_{\rm IR}^{(1)}=0\,,\label{eq:boxa}
\\
&\hat{\Box} p+4 p-32(1+3h_0)(1+2h_0)a+\frac{64}{3} C_{\rm IR}^{(1)}\left(1+3h_0\right)^2=0\,,\label{eq:boxp}
\end{align}
where $\hat{\Box}\equiv D_{\hat{I}}D^{\hat{I}}$ is the Laplacian operator on $S^3$, which in $(x,y,\phi)$ coordinates reads
\begin{equation}
\hat{\Box} f\equiv \frac{1}{\sqrt{1-x^2}}\partial_x\left[(1-x^2)^{3/2}\partial_x f\right]+\frac{1}{1-x^2}\left\{\partial_y[(1-y^2)\partial_y f]+\frac{1}{1-y^2}\partial^2_{\phi}f\right\}\,,
\end{equation}
\end{subequations}
for some function $f$ on $S^3$. Let us focus for a moment on Eq.~(\ref{eq:boxa}). First, we note that we can easily remove the non-homogeneous term via a shift of the form
\begin{equation}
a=\tilde{a}+\frac{4}{3}\frac{1+3\,h_0}{1+h_0}C_{\rm IR}^{(1)}\,,
\end{equation}
which brings Eq.~(\ref{eq:boxa}) to
\begin{equation}
\hat{\Box} \tilde{a}+8\left(1+h_0\right)\tilde{a}=0\,.
\label{eq:atilde}
\end{equation}
We can now expand $\tilde{a}$ (and thus $a$) as a \emph{sum} of spherical harmonics $Y_{\ell\,k\,m}$ with a given value of $\ell$ (with a total of $(\ell+1)^2$ terms in such sum): 
\begin{equation}
\tilde a =\sum_{k=0}^{\ell}\sum_{m=-k}^k\,b^{\ell\,k\,m}Y_{\ell\,k\,m}(x,y,\phi)
 \,,
\label{eq:deformtot}
\end{equation}
which brings Eq.~(\ref{eq:atilde}) to
\begin{equation}
h_0=h_0^{\ell}\equiv \frac{1}{8}(\ell-2)(\ell+4)\,.
\label{eq:recast}
\end{equation}
The above is just a restatement of Eq.~(\ref{eq:yc}). At this stage, the coefficients $b^{\ell\,k\,m}$ are left \emph{undetermined}. Once we know $\tilde{a}$ it is not hard to find $p$ from (\ref{eq:boxp}) which yields 
\begin{equation}
p=16\,C^{(1)}_{\rm IR}\,(1+3h_0^{\ell})^2-8(1+3h_0^{\ell})a\,.
\label{eq:deformp}
\end{equation}
  Let us recap what we have achieved so far. We found that, for a particular value of $\ell$, first order smooth deformations of $AdS_2\times S^3$ given by (\ref{eq:deformtot}) and (\ref{eq:deformp}) exist, so long as $h_0$ is given by Eq.~(\ref{eq:recast}).

We will only need the equation for $h_1^{(2)}\equiv s$ to break the degeneracy, \emph{i.e.} to find the $b^{\ell\,k\,m}$. The equation for this particular function can be easily obtained by expanding the $tt$ component of the Einstein-DeTurck equation to second order in $\varepsilon$, which yields
\begin{subequations}
\begin{equation}
\hat{\Box}s+8(1+h_0^{\ell})s=T\,.
\label{eq:equations}
\end{equation}
with 
\begin{multline}
T\equiv
(1+3h_0^{\ell})\Bigg[\frac{32}{3}C_{\rm IR}^{(2)}+\frac{512(1+3h_0^{\ell})^2}{9(1+h_0^{\ell})^2}{C_{\rm IR}^{(1)}}^2
+\frac{32}{h_0^{\ell}}(3+4h_0^{\ell})\tilde{a}^2
\\
+\frac{4}{h_0^{\ell}}\hat{g}^{\hat{I}\hat{J}}\hat{D}_{\hat{I}}\tilde{a}\hat{D}_{\hat{J}}\tilde{a}-\frac{128(1+3h_0^{\ell})^2}{3(1+h_0^{\ell})}\,C_{\rm IR}^{(1)}\,\tilde{a}\Bigg]\,.
\end{multline}
\end{subequations}

The source term $T$ in Eq.~(\ref{eq:equations}) behaves as a scalar on $S^3$ and is quadratic in $\tilde a$.  Since we are assuming that $\tilde a$ only includes  harmonics of fixed $\ell$, $T$ only includes harmonics up to $2\ell$. Furthermore, since $T$ is a smooth function on $S^3$, it can also be decomposed as a sum of harmonics, and thus we can set
\begin{equation}
T=\sum_{\tilde{\ell}=0}^{2\ell}\sum_{k=0}^{\tilde{\ell}}\sum_{m=-k}^{k}T^{\tilde{\ell}\,k\,m}Y_{\tilde{\ell}\,k\,m}\,.
\label{eq:sumT}
\end{equation}

Smooth solutions to Eq.~(\ref{eq:equations}) exist only if the sum (\ref{eq:sumT}) has no component with $\tilde{\ell}=\ell$, otherwise there is a resonance in Eq.~(\ref{eq:equations}) and the solutions are necessarily singular. Assuming that no such component exists, the smooth solution to Eq.~(\ref{eq:equations}) can be written as
\begin{equation}
s=\sum_{k=0}^{\ell}\sum_{m=-k}^k\,\tilde{b}^{\ell\,k\,m}Y_{\ell\,k\,m}+\sum_{\tilde{\ell}=0}^{2\ell}\sum_{k=0}^{\tilde{\ell}}\sum_{m=-k}^{k}\frac{T^{\tilde{\ell}\,k\,m}}{\tilde{\ell}(\tilde{\ell}+2)-8(1+h_0^{\ell})}Y_{\tilde{\ell}\,k\,m}\,,
\label{eq:secondgen}
\end{equation}
where the first term is the solution to the homogeneous equation and $\tilde{b}^{\ell\,k\,m}$ are undetermined constants. Thus the existence of smooth solutions boils down to the problem of determining whether we can choose $b^{\ell\,k\,m}$ of the first order solution, so that no terms with $\tilde{\ell}=\ell$ appear in the sum (\ref{eq:sumT}). This requirement translates into quadratic constraints amongst the $b^{\ell\,k\,m}$ and $C_{\rm IR}^{(1)}$, which might or might not admit real solutions. Note that the procedure outlined above also works at higher orders, and indeed, the homogeneous term in (\ref{eq:secondgen}) is fixed at higher order, though the higher order constraints are linear in $\tilde{b}^{\ell\,k\,m}$ so long as there are no further degeneracies amongst the first order $b^{\ell\,k\,m}$. As such, smooth solutions are guaranteed to exist so long as we can determine real solutions to $b^{\ell\,k\,m}$ and $C_{\rm IR}^{(1)}$, and there are no further degeneracies left in $b^{\ell\,k\,m}$.

The constraints that we want to impose amount to:
\begin{equation}
\int_0^{2\pi}\int_{-1}^1\int_{-1}^1\,\sqrt{1-x^2}\,T\,Y_{\ell\,k\,m}(x,y,\phi)\,\mathrm{d}\phi\,\mathrm{d}x\,\mathrm{d}y=0\,,
\label{eq:condition}
\end{equation}
where we regard $\ell$ as given, but $k$ and $m$ take the appropriate ranges. This means that there are a total of $(\ell+1)^2$ constraints for a given value of $\ell$. This is precisely the number of $b^{\ell\,k\,m}$ variables that we have at our disposal and one might worry that the only possible solution to the constraint above has $b^{\ell\,k\,m}=0$. However, we recall that the constraint will in general also depend on $C^{(1)}_{\rm IR}$, which should also be determined in this procedure. Let us attempt to evaluate the above integral. Since we are taking $\ell\geq3$, the first two terms in $T$ do not contribute. Also, the last term will simply yield {a term proportional to} $b^{\ell\,k\,m}$. 

Let us define the following overlap integrals\footnote{Explicit expressions for $A_{\ell\,k\,m;\tilde{\ell}\,\tilde{k}\,\tilde{m};\hat{\ell}\,\hat{k}\,\hat{m}}$ can be found using the results of  \cite{2005math.ph...9035A}.}
\begin{subequations}
\begin{align}
A_{\ell\,k\,m;\tilde{\ell}\,\tilde{k}\,\tilde{m};\hat{\ell}\,\hat{k}\,\hat{m}}&=\int_0^{2\pi}\int_{-1}^{1}\int_{-1}^1 \sqrt{1-x^2}\;Y_{\ell\,k\,m}\;Y_{\tilde{\ell}\,\tilde{k} \tilde{m}}\;Y_{\hat{\ell}\,\hat{k} \hat{m}}\,\mathrm{d}\phi\,\mathrm{d}x\,\mathrm{d}y\,,
\\
B_{\ell\,k\,m;\tilde{\ell}\,\tilde{k}\,\tilde{m};\hat{\ell}\,\hat{k}\,\hat{m}}&=\int_0^{2\pi}\int_{-1}^{1}\int_{-1}^1 \sqrt{1-x^2}\;Y_{\ell\,k\,m}\;\hat{\nabla}^{\hat{I}}Y_{\tilde{\ell}\,\tilde{k} \tilde{m}}\;\hat{\nabla}_{\hat{I}}Y_{\hat{\ell}\,\hat{k} \hat{m}}\,\mathrm{d}\phi\,\mathrm{d}x\,\mathrm{d}y\nonumber
\\
& = \frac{1}{2}\left[\tilde{\ell}(\tilde{\ell}+2)+\hat{\ell}(\hat{\ell}+2)-\ell(\ell+2)\right]A_{\ell\,k\,m;\tilde{\ell}\,\tilde{k}\,\tilde{m};\hat{\ell}\,\hat{k}\,\hat{m}}\,.
\end{align}
\end{subequations}
For odd values of $\ell$, it is a simple exercise to show that both $A_{\ell\,k\,m;\tilde{\ell}\,\tilde{k}\,\tilde{m};\hat{\ell}\,\hat{k}\,\hat{m}}$ and $B_{\ell\,k\,m;\tilde{\ell}\,\tilde{k}\,\tilde{m};\hat{\ell}\,\hat{k}\,\hat{m}}$ vanish identically, due to parity considerations. In such a case, $C^{(1)}_{\rm IR}=0$ and one has to go one order higher to find all the $b^{\ell\,k\,m}$. We shall focus on even values of $\ell$ hereafter.

In terms of the symbols above, the condition (\ref{eq:condition}) translates into
\begin{multline}
\frac{3(1+h_0^{\ell})}{32\,h_0^{\ell}\,(1+3h_0^{\ell})^2}\Bigg[8(3+4h_0^{\ell})\sum_{\tilde{k}=0}^{\ell}\sum_{\tilde{m}=-\tilde{k}}^{\tilde{k}}\sum_{\hat{k}=0}^{\ell}\sum_{\hat{m}=-\hat{k}}^{\hat{k}}A_{\ell\,k\,m;\ell\,\tilde{k}\,\tilde{m};\ell\,\hat{k}\,\hat{m}}b^{\ell\,\tilde{k}\,\tilde{m}}b^{\ell\,\hat{k}\,\hat{m}}
\\
+\sum_{\tilde{k}=0}^{\ell}\sum_{\tilde{m}=-\tilde{k}}^{\tilde{k}}\sum_{\hat{k}=0}^{\ell}\sum_{\hat{m}=-\hat{k}}^{\hat{k}}B_{\ell\,k\,m;\ell\,\tilde{k}\,\tilde{m};\ell\,\hat{k}\,\hat{m}}b^{\ell\,\tilde{k}\,\tilde{m}}b^{\ell\,\hat{k}\,\hat{m}}
\Bigg]=C^{(1)}_{\rm IR}b^{\ell\,k\,m}\,.
\label{eq:crazynuts}
\end{multline}

Let us define the $(\ell+1)^2$-dimensional vector $\bf{X}$ with
\begin{equation}
\mathbf{X}=\{b^{\ell\,0\,0},b^{\ell\,1\,-1},b^{\ell\,1\,0},b^{\ell\,1\,1},b^{\ell\,2\,-2},b^{\ell\,2\,1},b^{\ell\,2\,0},b^{\ell\,2\,1},b^{\ell\,2\,2}\,\ldots,b^{\ell\,\ell\,\ell-1},b^{\ell\,\ell\,\ell}\}\,,
\end{equation}
in terms of which we can recast (\ref{eq:crazynuts}) in the following form 
\begin{subequations}
\label{eq:zeigenvalue}
\begin{equation}
{K^{\mathfrak{p}}_{\mathfrak{m}\mathfrak{n}}}X^{\mathfrak{m}}X^{\mathfrak{n}}=C^{(1)}_{\rm IR}\,X^{\mathfrak{p}}
\label{eq:zeigenvaluea}
\end{equation}
for an appropriate choice of tensor ${K^{\mathfrak{p}}_{\mathfrak{m}\mathfrak{n}}}$, with Gothic indices being $(\ell+1)^2$ dimensional and using the Einstein summation convention. Without a choice of normalisation, which ultimately fixes $\varepsilon$, (\ref{eq:zeigenvaluea}) admits an infinite number of solutions. We thus demand
\begin{equation}
\delta_{\mathfrak{m}\mathfrak{n}} X^{\mathfrak{m}}X^{\mathfrak{n}}=1\,.
\end{equation}
\end{subequations}

The system of algebraic equations (\ref{eq:zeigenvalue}) is often referred to in the literature as a  $Z-$eigenvalue problem. The expansion parameter $C^{(1)}_{\rm IR}$ plays the role of the $Z-$eigenvalue. Note that just from the structure of the $Z-$eigenvalue, we can conclude that if $\{{\bf X},C^{(1)}_{\rm IR}\}$ is a $Z-$eigenpair, so is  $\{-{\bf X},-C^{(1)}_{\rm IR}\}$. For this reason, we will focus on values of $C^{(1)}_{\rm IR}$ that are positive. 

Unfortunately, unlike for standard eigenvalue problems, there are not many numerical methods available to determine \emph{all} $Z-$eigenvalues. We thus have to proceed via a standard Newton-Raphson algorithm.

We were able to find some particular analytic solutions to the above $Z-$eigenvalue problem when insisting on preserving $SO(3)$ (recovering the results of section \ref{sec:ell2}) or $U(1)$. However, the main interest of this section is to show that solutions exist that break all rotational symmetries. We thus proceed numerically for particular values of $\ell$, by randomizing our initial seeds and running a standard Newton-Raphson algorithm. In total, we found $4$, $3$ and $49$ non-trivial values of $C^{(1)}_{\rm IR}$ for $\ell=4,6,8$, respectively.

So far we have detailed a method for constructing a particular configuration \{$b^{\ell\,k\,m},C^{(1)}_{\rm IR}\}$. However, it could be that such a configuration would still preserve some rotational symmetries. To see that this was not the case, we computed $\pounds_{\Xi}a$  with $\Xi = \sum_{i=1}^6 u^{(i)} \xi^{(i)}$, where $\xi^{(i)}$ are the six rotational Killing fields of the round three-sphere and $u^{(i)}$ are some constants. If we find configurations for which $\pounds_{\Xi}a=0$ implies $u^{(i)}=0$, such configurations necessarily break all rotational symmetries. Note that this statement depends on the values of $b^{\ell\,k\,m}$ we obtain by solving our $Z-$ eigenvalue problem. Indeed, we found some configurations which did preserve a subset of rotational symmetries of the original $S^3$. However, for the values quoted in table \ref{tab:1}, we explcitly checked that all rotational symmetries are broken.
\begin{table}
\centering
\begin{tabular}{|l|l|l|}
\hline\hline
\multicolumn{3}{|c|}{$C^{(1)}_{\rm IR}$}
\\
\hline\hline
\multicolumn{1}{|c|}{$\ell=4$} & \multicolumn{1}{c}{$\ell=6$} & \multicolumn{1}{|c|}{$\ell=8$}
\\
\hline\hline
$0.0184558$ & $0.0046848$ & $7.83170\times 10^{-7}$
\\
$0.0583625$ & $0.0205736$ & $0.0000132116$
\\
\hline\hline
\end{tabular}
\caption{\label{tab:1}The smallest two numerical values for $C^{(1)}_{\rm IR}$ found for several values of $\ell$. All values displayed in this table correspond to configurations that break \emph{all} rotational symmetries.}
\end{table}

We also solved our problem fully nonlinearly using the numerical methods of \cite{Santos:2023} and found perfect agreement with the linear calculations above. Of course, nonlinearly, one can do better and for instance predict what $C^{(2)}_{\rm IR}$ should be for a given value of $\ell$. For a $\ell=4$ configuration with $C^{(1)}_{\rm IR}=0.0184558$, our fully nonlinear calculation predicts $C^{(2)}_{\rm IR}=-0.6321496$, where at the nonlinear level we defined $\varepsilon$ as
\begin{subequations}
\begin{equation}
Z_{km}\equiv \int_0^{2\pi}\int_{-1}^{1}\int_{-1}^1 \sqrt{1-x^2}\;Y_{4\,k\,m}(x,y,\phi)\;H_1(x,y,\phi)\,\mathrm{d}\phi\,\mathrm{d}x\,\mathrm{d}y\,,
\end{equation}
with
\begin{equation}
\sum_{k=0}^{4}\sum_{m=-k}^{k}Z_{km}Z_{km}=\varepsilon^2\,.
\end{equation}
\end{subequations}
\section{Discussion}

We have seen that the extremal Reissner-Nordstr\"om AdS solution does not provide a good dual description of the generic IR behavior of four (or higher) dimensional holographic theories. This is because its near horizon geometry, $AdS_2\times S^3$, is unstable to static perturbations that break $SO(4)$. We have constructed a new family of near horizon geometries, labelled by the charge $Q$, and shown that they are stable to $SO(3)$-invariant linearized perturbations. Moreover, they are stable to nonlinear perturbations in this class, since they arise in the $T\to 0$ limit of an open family of $SO(3)$-invariant AdS black holes. Thus, following the usual holographic dictionary, under this reduced symmetry they represent stable IR fixed points of a dual  RG flow.

Although our new IR geometries have the property that perturbations go to zero at the horizon, they are not {generically} completely smooth. As shown in Fig.~\ref{fig:per}, they go to zero like a power law with a power $\gamma$ that is much less than one. This means that if we take two derivatives to compute the curvature, certain components will diverge. In other words, infalling observers experience diverging tidal forces at the horizon. For the solutions constructed in Sec. 4 that approach our new IR geometries, we have computed certain components of the Weyl tensor on the horizon as a function of $T$.  We find that they diverge as $T\to 0$ in a way consistent with the perturbative argument in Sec. 3.3.

This is exactly analogous to the singularities found in four-dimensional extremal black holes \cite{Horowitz:2022mly}. The main difference is that in four bulk dimensions, the horizon geometry remains $AdS_2\times S^2$ and does not get distorted. As shown in  \cite{Horowitz:2022mly}, all curvature scalars remain finite at the horizon, so if one analytically continues the solution to obtain a Euclidean black hole, it is completely smooth. The same will be true for the five-dimensional solutions constructed here. These singularities are only a feature of the Lorentzian solution, although they can affect thermodynamic quantities like the specific heat.

 Supersymmetric black holes in $AdS_5$ exist which are rotating and have smooth horizons \cite {Gutowski:2004yv}. It would be interesting to study how their near horizon geometry responds to small deformations of the boundary conditions. Given the results of this paper and the fact that these deformations break supersymmetry, we expect the near horizon geometry will be significantly altered.

An important open problem is to find the generic stable IR geometry. In Appendix A we construct another large class of  $SO(3)$-invariant near horizon geometries which are associated with $\ell > 2$ instabilities of $AdS_2\times S^3$ that only arise at large enough $Q$. However, all of them have at least one unstable $SO(3)$-invariant mode, so they are not stable RG fixed points, even under this reduced symmetry. We have also shown how to construct a very large class of near horizon geometries without any rotational symmetry in Sec. 5. These solutions exist close to $AdS_2\times S^3$ when the $S^3$ is large enough. However we expect these solutions will also  be unstable since the unstable mode of $AdS_2\times S^3$ should persist for the new solutions, via continuity.

We have restricted our attention to black holes in global AdS, dual to holographic theories on $S^3 \times \mathbb{R}$. The generic IR behavior of theories on $\mathbb{R}^4$ or $T^3\times \mathbb{R}$ is equally interesting and under investigation. We hope to report our results soon.

\subsection*{Acknowledgments}
It is a pleasure to thank Jan~Boruch, Sean~Hartnoll, and Harvey~Reall for discussions. The work of G.~H. was supported in part by NSF Grant PHY-2107939. J.~E.~S. has been partially supported by STFC consolidated grant ST/T000694/1. M.~K. thanks University of California in Santa Barbara for the hospitality, he was partially supported by The Polish-U.S. Fulbright Commission.

\appendix

\section{$SO(3)$-invariant near horizon geometries associated with $\ell >2$ modes}

It is clear from Eq.~(\ref{eq:yc}) that for each value of $\ell>2$, there are novel near horizon geometries that connect smoothly to $AdS_2\times S^3$. These occur precisely at $y_+=y_+^c(\ell)$. In this appendix, we construct $SO(3)$-invariant solutions in this class both perturbatively and nonperturbatively. Unlike the solutions discussed in Sec. 3, these solutions all remain unstable to an $SO(3)$-invariant $\ell =2$ perturbation.

 It turns out that the new solutions associated with even values of $\ell$ and odd values of $\ell$ behave very differently. For even values of $\ell$, and within our symmetry assumptions, there are exactly two solutions emerging from $y_+=y_+^c(\ell)$, whereas for odd values of $\ell$ there exists a single family. The reason for this is that for odd $\ell$ we can change the sign of the perturbation direction, \emph{i.e.} of $\varepsilon$, by doing a reflection around the equatorial plane $\theta\to\pi-\theta$, which is a symmetry of the Bertotti-Robinson background solution. However, this is not the case for even values of $\ell$. For this reason, all physical observables for odd values of $\ell$ can only depend on $\varepsilon^2$.

\subsection{\label{app:expansion}Perturbative expansion for particular values of $\ell$ up to $\varepsilon^6$}

The perturbation scheme is slightly different from that presented in Eq.~(\ref{eq:per}). The main difference being that we are now expanding about a non-singular solution, and as such $B$, $\rho^2_{\rm IR}$ and $Y_+^2$ all have order $\varepsilon^0$ terms. Also, our perturbation parameter $\varepsilon$, is defined as
\begin{equation}
\int_0^{\pi}B \sin^2\theta\,Y_{\ell}(\theta)\,\mathrm{d}\theta=\varepsilon\,.
\end{equation}
This normalisation is not possible for the $\ell=2$ case, because $B$ itself starts at order $\varepsilon$ in that case. Below we list the results of our perturbations scheme for $\ell=3,4,5,6,7,8$ up to $\mathcal{O}(\varepsilon^6)$.
\subsection*{ $\ell=3$:}
\begin{subequations}
\begin{multline}
\frac{Q(\varepsilon)}{L^2}=\frac{7 \sqrt{33} \pi }{32}+\frac{368054399}{6468 \sqrt{33}} \varepsilon ^2+\frac{411220142620386290969}{135898460775 \sqrt{33} \pi } \varepsilon^4
\\
+\frac{1296275883356827283869527398373715393 }{5048255476393495616250 \sqrt{33} \pi ^2}\varepsilon^6+O\left(\varepsilon^7\right)
\end{multline}
\begin{multline}
\frac{\Delta S(\varepsilon)}{L^2} =-\frac{736108798 }{1617} \sqrt{\frac{2}{7}}\varepsilon^4-\frac{19564147914807179502992 }{407695382325 \pi } \sqrt{\frac{2}{7}}\varepsilon^6+O\left(\varepsilon^7\right)
\end{multline}
\end{subequations}
\vskip .2cm
\subsection*{$\ell=4$:}
\begin{subequations}
\begin{multline}
\frac{Q(\varepsilon)}{L^2}=\sqrt{15} \pi +245 \sqrt{\frac{15 \pi }{2}} \varepsilon +\frac{140581}{2} \sqrt{\frac{5}{3}} \varepsilon ^2+\frac{10156181}{2} \sqrt{\frac{15}{2 \pi }} \varepsilon ^3+\frac{37986958009 }{36
   \pi } \sqrt{\frac{5}{3}}\varepsilon^4
   \\
   +\frac{16465994398343}{54 \pi ^{3/2}}  \sqrt{\frac{5}{6}}\varepsilon^5+\frac{4572303927656731}{216 \sqrt{15} \pi ^2} \varepsilon ^6+O\left(\varepsilon^7\right)
\end{multline}
\begin{multline}
\frac{\Delta S(\varepsilon)}{L^2} =-9800 \sqrt{\pi } \varepsilon ^3-1376410 \sqrt{2} \varepsilon^4-\frac{2594844637}{6 \sqrt{\pi }} \varepsilon^5-\frac{6593328568855}{108 \sqrt{2} \pi } \varepsilon^6+O\left(\varepsilon^7\right)
\end{multline}
\end{subequations}
\subsection*{$\ell=5$:}
\begin{subequations}
\begin{multline}
\frac{Q(\varepsilon)}{L^2}=\frac{27 \sqrt{93} \pi }{32}+\frac{24400434260923 }{24538140 \sqrt{93}}\varepsilon^2+\frac{2117065639783090195016175652951 }{17670613036913612325 \sqrt{93} \pi
   }\varepsilon^4
   \\
   +\frac{72052767378571194925798463365202395227894683041956279121}{3306124766831924422071634047639626587500 \sqrt{93} \pi ^2} \varepsilon^6+O\left(\varepsilon^7\right)
\end{multline}
\begin{multline}
\frac{\Delta S(\varepsilon)}{L^2} =-\frac{48800868521846 }{18403605} \sqrt{\frac{2}{3}}\varepsilon^4-\frac{15837483245656122360043297709584  }{25650889892293953375 \pi }\sqrt{\frac{2}{3}}\varepsilon^6+O\left(\varepsilon^7\right)
\end{multline}
\end{subequations}
\subsection*{$\ell=6$:}
\begin{subequations}
\begin{multline}
\frac{Q(\varepsilon)}{L^2}=\frac{5 \sqrt{33} \pi }{2}+\frac{13312}{5} \sqrt{\frac{6 \pi }{11}} \varepsilon +\frac{9210093568}{4125 \sqrt{33}} \varepsilon^2+\frac{1059863200006144 }{3403125} \sqrt{\frac{2}{33 \pi }}\varepsilon^3
\\
+\frac{194345221205728267599872 }{2299406484375 \sqrt{33} \pi }\varepsilon^4+\frac{14996706341255368171002527744 }{1138206209765625 \pi^{3/2}}\sqrt{\frac{2}{33}}\varepsilon^5
   \\
   +\frac{3704537017903529074897132775725334528 }{1281762467972314453125 \sqrt{33} \pi ^2}\varepsilon^6+O\left(\varepsilon^7\right)
\end{multline}
\begin{multline}
\frac{\Delta S(\varepsilon)}{L^2} =-\frac{212992}{5} \sqrt{\frac{2 \pi }{5}} \varepsilon ^3-\frac{6618087424}{375 \sqrt{5}} \varepsilon^4-\frac{323703625744384 }{84375} \sqrt{\frac{2}{5 \pi }}\varepsilon^5
\\
-\frac{64632743045075027623936}{57010078125 \left(\sqrt{5} \pi \right)} \varepsilon^6+O\left(\varepsilon^7\right)
\end{multline}
\end{subequations}
\subsection*{$\ell=7$:}
\begin{subequations}
\begin{multline}
\frac{Q(\varepsilon)}{L^2}=\frac{55 \sqrt{177} \pi }{32}+\frac{49470716440836139 }{23679155500} \sqrt{\frac{3}{59}}\varepsilon^2
\\
+\frac{674695516051872003625594714069783588619}{485541698594507874841396875 \sqrt{177} \pi
   } \varepsilon^4
   \\
   +\frac{1176498595333690205420678287619561589933772224186864713714038571814170461 }{2548337217860698436964062929388348722320239754199218750 \sqrt{177} \pi ^2}\varepsilon^6
   \\
   +O\left(\varepsilon^7\right)
\end{multline}
\begin{multline}
\frac{\Delta S(\varepsilon)}{L^2} =-\frac{296824298645016834 }{5919788875} \sqrt{\frac{2}{55}}\varepsilon^4
\\
-\frac{13290175140845121928392418684118371107376  }{617214023637086281578046875 \pi }\sqrt{\frac{2}{55}}\varepsilon^6+O\left(\varepsilon ^7\right)
\end{multline}
\end{subequations}
\subsection*{$\ell=8$:}
\begin{subequations}
\begin{multline}
\frac{Q(\varepsilon)}{L^2}=\frac{9 \sqrt{57} \pi }{2}+\frac{68992}{3} \sqrt{\frac{2 \pi }{57}} \varepsilon +\frac{33256552448}{3705 \sqrt{57}}\varepsilon ^2+\frac{4867733289052168192}{3335667075} \sqrt{\frac{2}{57 \pi }} \varepsilon^3
\\
+\frac{74581693749397055463423311872}{112317851238310575 \sqrt{57} \pi } \varepsilon^4
\\
+\frac{35579600837189584397112696595360514048 }{318345293711024620486875 \pi
   ^{3/2}}\sqrt{\frac{2}{57}} \varepsilon^5
   \\
   +\frac{8932378258346660177296551392555630719690487103488}{160788795183274474599632525615625 \sqrt{57} \pi ^2} \varepsilon^6+O\left(\varepsilon ^7\right)
\end{multline}
\begin{multline}
\frac{\Delta S(\varepsilon)}{L^2} =-\frac{1103872}{27} \sqrt{2 \pi } \varepsilon^3-\frac{124997648384}{5265} \varepsilon^4-\frac{173099850617716736 }{27720225}\sqrt{\frac{2}{\pi }} \varepsilon^5
\\
-\frac{51087108987737527073374208}{16375251674925 \pi } \varepsilon^6+O\left(\varepsilon ^7\right)
\end{multline}
\end{subequations}

\subsection{Nonlinear solutions}

\begin{figure}[th]
\centering \includegraphics[width=0.9\textwidth]{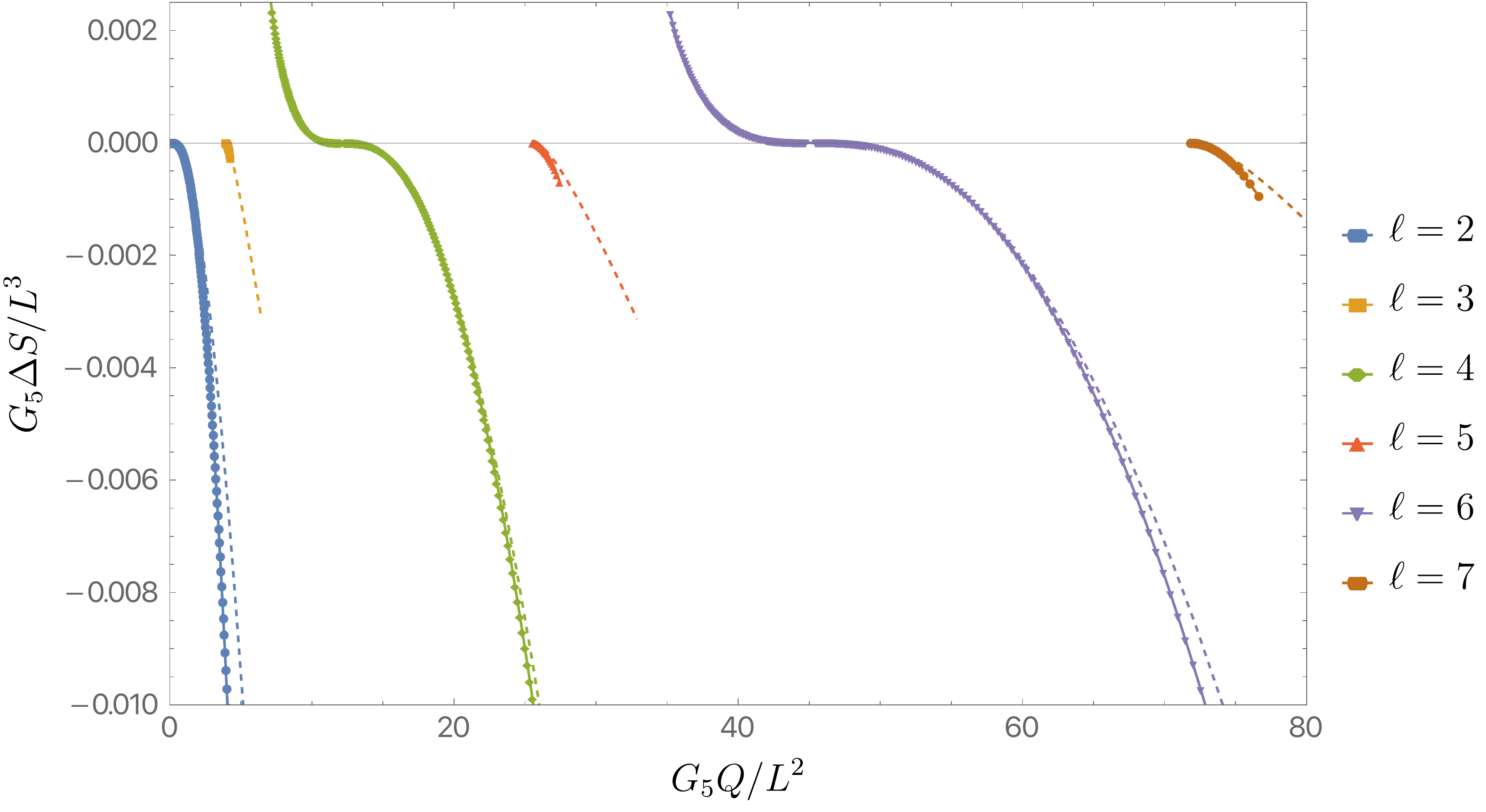}
\caption{\label{fig:entropy_several} $\Delta S$ as a function of $Q$ for the new IR geometries near their onset, for $\ell=2,\ldots,7$. The several dashed lines give the perturbative expansions detailed in \ref{sec:ell2} and above, while the several coloured symbols how our exact numerical data. The several values of $\ell$ are distinguished by colour, and are labelled on the figure.}
\end{figure}

We have also constructed the full nonlinear solutions in the same way that we constructed the solutions in Sec.~3.2.
In Fig.~\ref{fig:entropy_several} we plot the difference in entropy between these new solutions and RN AdS (with the same charge) as a function of $Q/L^2$ for $\ell=2,\ldots,7$. The dashed coloured lines represent the perturbative expansion detailed above, while the several symbols show the numerical data extracted non-linearly using our numerical scheme. The colour coding used is indicated in the legend. Our numerical scheme and perturbative results agree well near the several onsets for $\ell\geq3$. We also include the $\ell =2$ solutions discussed in the body of the paper  near $Q=0$ for comparison. Unlike the $\ell=2$ solutions, these new solutions do not extend to all $Q$, but instead appear to become singular eventually. In fact, the odd $\ell$ branches of solutions do not extend much beyond their onset.

We have studied the RG stability of these higher $\ell$ near horizon geometries, following the procedure described in Sec.~\ref{RGstable}.  We find that they remain RG unstable to $\ell=2$ perturbations, although the scaling exponent $\gamma$ is not very negative.

\section{\label{app:crazy_long}Expressions for the equations governing the IR perturbations in Sec.~\ref{RGstable}}
\begin{subequations}
\begin{equation}
\alpha_0 \equiv \frac{8 \rho _{\text{IR}}^2 Y_+^2H^2}{9 B^2 B'} \left[3 B (1-4 B)-4 \rho _{\text{IR}}^2\right]\,,
\end{equation}
\begin{multline}
\alpha_1 \equiv\frac{8 \rho _{\text{IR}}^2 Y_+^2 \sin \theta H^3}{9 B^2 B'} \Big\{4 Y_+^2 \left[3 B(1-4 B)-4 \rho _{\text{IR}}^2\right]
   \left[\rho _{\text{IR}}^2-B (1-6 B)\right] H^2
   \\
   -9 B (1-12 B) {B'}^2\Big\}+ \frac{48 \rho _{\text{IR}}^2 H^4 B'}{\sin\theta}\,,
\end{multline}
\begin{equation}
\alpha_2\equiv \frac{16 \rho _{\text{IR}}^2 Y_+^2 \sin \theta H^3}{9 BB'} \left\{2 Y_+^2 \left[4 \rho _{\text{IR}}^2-3 B(1-4 B)\right] H^2+9 {B'}^2\right\}\,,
\end{equation}
\begin{equation}
\beta_0 \equiv \frac{4 Y_+^2 H^2}{3 B^2} \left(7 \rho _{\text{IR}}^2-3 B\right)
\end{equation}
\begin{equation}
\beta_1 \equiv -\frac{2 Y_+^2 \left(2 \rho _{\text{IR}}^2-B\right) \sin\theta H^3}{3 B^2} \left[4 Y_+^2 \left(5 \rho _{\text{IR}}^2-3 B+6 B^2\right) H^2+9 {B'}^2\right]\,,
\end{equation}
\begin{multline}
\beta_2 \equiv \frac{H^2 Y_+^2}{3 B} \Big\{9 H B' \left(8 B \cos\theta+\sin\theta B'\right)-36 B \sin \theta B' H'
\\
-4 H^3 \sin\theta\,Y_+^2 \left[23 \rho _{\text{IR}}^2-3 B (5-14 B)\right]\Big\}\,,
\end{multline}
\begin{equation}
\beta_3 \equiv 4 Y_+^4 \sin \theta H^5\,,
\end{equation}
\begin{equation}
\kappa_0\equiv -\frac{28 \rho _{\text{IR}}^2 Y_+^2 H^2}{3 B^2}\,,
\end{equation}
\begin{equation}
\kappa_1 \equiv \frac{4 \rho _{\text{IR}}^2 Y_+^2 \sin\theta H^3}{3 B^2} \left\{4 Y_+^2 \left[5 \rho _{\text{IR}}^2-3 B(1-2 B)\right] H^2+9 {B'}^2\right\}\,,
\end{equation}
and
\begin{equation}
\kappa_2 = \frac{16 \rho _{\text{IR}}^2 Y_+^4 \sin \theta H^5}{B}\,.
\end{equation}
\end{subequations}
\bibliographystyle{JHEP}
\bibliography{new_ir}

\providecommand{\href}[2]{#2}\begingroup\raggedright\begin{thebibliography}{10}

\bibitem{Hartnoll:2011fn}
S.~A. Hartnoll, \emph{{Horizons, holography and condensed matter}},  in
  \emph{{Black holes in higher dimensions}} (G.~T. Horowitz, ed.), (Cambridge,
  UK), pp.~387--419, Cambridge Univ. Pr., 2012.
\newblock \href{https://arxiv.org/abs/1106.4324}{{\ttfamily 1106.4324}}.

\bibitem{Peet:1998wn}
A.~W. Peet and J.~Polchinski, \emph{{UV / IR relations in AdS dynamics}},
  \href{http://dx.doi.org/10.1103/PhysRevD.59.065011}{\emph{Phys. Rev. D}
  {\bfseries 59} (1999) 065011},
  [\href{https://arxiv.org/abs/hep-th/9809022}{{\ttfamily hep-th/9809022}}].

\bibitem{Dias:2021vve}
O.~J.~C. Dias, G.~T. Horowitz and J.~E. Santos, \emph{{Extremal black holes
  that are not extremal: maximal warm holes}},
  \href{http://dx.doi.org/10.1007/JHEP01(2022)064}{\emph{JHEP} {\bfseries 01}
  (2022) 064}, [\href{https://arxiv.org/abs/2109.14633}{{\ttfamily
  2109.14633}}].

\bibitem{Horowitz:2022mly}
G.~T. Horowitz, M.~Kolanowski and J.~E. Santos, \emph{{Almost all extremal
  black holes in AdS are singular}},
  \href{https://arxiv.org/abs/2210.02473}{{\ttfamily 2210.02473}}.

\bibitem{Kunduri:2013gce}
H.~K. Kunduri and J.~Lucietti, \emph{{Classification of near-horizon geometries
  of extremal black holes}},
  \href{http://dx.doi.org/10.12942/lrr-2013-8}{\emph{Living Rev. Rel.}
  {\bfseries 16} (2013) 8}, [\href{https://arxiv.org/abs/1306.2517}{{\ttfamily
  1306.2517}}].

\bibitem{DHoker:2002nbb}
E.~D'Hoker and D.~Z. Freedman, \emph{{Supersymmetric gauge theories and the AdS
  / CFT correspondence}},  in \emph{{Theoretical Advanced Study Institute in
  Elementary Particle Physics (TASI 2001): Strings, Branes and EXTRA
  Dimensions}}, pp.~3--158, 1, 2002.
\newblock \href{https://arxiv.org/abs/hep-th/0201253}{{\ttfamily
  hep-th/0201253}}.

\bibitem{Dias:2015nua}
O.~J.~C. Dias, J.~E. Santos and B.~Way, \emph{{Numerical Methods for Finding
  Stationary Gravitational Solutions}},
  \href{http://dx.doi.org/10.1088/0264-9381/33/13/133001}{\emph{Class. Quant.
  Grav.} {\bfseries 33} (2016) 133001},
  [\href{https://arxiv.org/abs/1510.02804}{{\ttfamily 1510.02804}}].

\bibitem{Smarr:1973zz}
L.~Smarr, \emph{{Surface Geometry of Charged Rotating Black Holes}},
  \href{http://dx.doi.org/10.1103/PhysRevD.7.289}{\emph{Phys. Rev. D}
  {\bfseries 7} (1973) 289--295}.

\bibitem{Costa:2015gol}
M.~S. Costa, L.~Greenspan, M.~Oliveira, J.~a. Penedones and J.~E. Santos,
  \emph{{Polarised Black Holes in AdS}},
  \href{http://dx.doi.org/10.1088/0264-9381/33/11/115011}{\emph{Class. Quant.
  Grav.} {\bfseries 33} (2016) 115011},
  [\href{https://arxiv.org/abs/1511.08505}{{\ttfamily 1511.08505}}].

\bibitem{Costa:2017tug}
M.~S. Costa, L.~Greenspan, J.~Penedones and J.~E. Santos, \emph{{Polarised
  Black Holes in ABJM}},
  \href{http://dx.doi.org/10.1007/JHEP06(2017)024}{\emph{JHEP} {\bfseries 06}
  (2017) 024}, [\href{https://arxiv.org/abs/1702.04353}{{\ttfamily
  1702.04353}}].

\bibitem{Headrick:2009pv}
M.~Headrick, S.~Kitchen and T.~Wiseman, \emph{{A New approach to static
  numerical relativity, and its application to Kaluza-Klein black holes}},
  \href{http://dx.doi.org/10.1088/0264-9381/27/3/035002}{\emph{Class.Quant.Grav.}
  {\bfseries 27} (2010) 035002},
  [\href{https://arxiv.org/abs/0905.1822}{{\ttfamily 0905.1822}}].

\bibitem{Adam:2011dn}
A.~Adam, S.~Kitchen and T.~Wiseman, \emph{{A numerical approach to finding
  general stationary vacuum black holes}},
  \href{http://dx.doi.org/10.1088/0264-9381/29/16/165002}{\emph{Class. Quant.
  Grav.} {\bfseries 29} (2012) 165002},
  [\href{https://arxiv.org/abs/1105.6347}{{\ttfamily 1105.6347}}].

\bibitem{Wiseman:2011by}
T.~Wiseman, \emph{{Numerical construction of static and stationary black
  holes}},  in \emph{{Black holes in higher dimensions}} (G.~T. Horowitz, ed.),
  pp.~233--270, 2012.
\newblock \href{https://arxiv.org/abs/1107.5513}{{\ttfamily 1107.5513}}.

\bibitem{Figueras:2011va}
P.~Figueras, J.~Lucietti and T.~Wiseman, \emph{{Ricci solitons, Ricci flow, and
  strongly coupled CFT in the Schwarzschild Unruh or Boulware vacua}},
  \href{http://dx.doi.org/10.1088/0264-9381/28/21/215018}{\emph{Class.Quant.Grav.}
  {\bfseries 28} (2011) 215018},
  [\href{https://arxiv.org/abs/1104.4489}{{\ttfamily 1104.4489}}].

\bibitem{Figueras:2016nmo}
P.~Figueras and T.~Wiseman, \emph{{On the existence of stationary Ricci
  solitons}}, \href{http://dx.doi.org/10.1088/1361-6382/aa764a}{\emph{Class.
  Quant. Grav.} {\bfseries 34} (2017) 145007},
  [\href{https://arxiv.org/abs/1610.06178}{{\ttfamily 1610.06178}}].

\bibitem{Horowitz:2012ky}
G.~T. Horowitz, J.~E. Santos and D.~Tong, \emph{{Optical Conductivity with
  Holographic Lattices}},
  \href{http://dx.doi.org/10.1007/JHEP07(2012)168}{\emph{JHEP} {\bfseries 07}
  (2012) 168}, [\href{https://arxiv.org/abs/1204.0519}{{\ttfamily 1204.0519}}].

\bibitem{Kunduri:2007vf}
H.~K. Kunduri, J.~Lucietti and H.~S. Reall, \emph{{Near-horizon symmetries of
  extremal black holes}},
  \href{http://dx.doi.org/10.1088/0264-9381/24/16/012}{\emph{Class. Quant.
  Grav.} {\bfseries 24} (2007) 4169--4190},
  [\href{https://arxiv.org/abs/0705.4214}{{\ttfamily 0705.4214}}].

\bibitem{Lindblom:2017maa}
L.~Lindblom, N.~W. Taylor and F.~Zhang, \emph{{Scalar, Vector and Tensor
  Harmonics on the Three-Sphere}},
  \href{http://dx.doi.org/10.1007/s10714-017-2303-y}{\emph{Gen. Rel. Grav.}
  {\bfseries 49} (2017) 139},
  [\href{https://arxiv.org/abs/1709.08020}{{\ttfamily 1709.08020}}].

\bibitem{Santos:2023}
J.~E. Santos, \emph{{Fast Spherical Transforms on the three-sphere}},
  \href{https://arxiv.org/abs/to appear}{{\ttfamily to appear}}.

\bibitem{2005math.ph...9035A}
S.~{Alisauskas}, \emph{{Integrals involving triplets of Jacobi and Gegenbauer
  polynomials and some 3j-symbols of SO(n), SU(n) and Sp(4)}}, {\emph{arXiv
  e-prints} (Sept., 2005) math--ph/0509035},
  [\href{https://arxiv.org/abs/math-ph/0509035}{{\ttfamily math-ph/0509035}}].

\bibitem{Gutowski:2004yv}
J.~B. Gutowski and H.~S. Reall, \emph{{General supersymmetric AdS(5) black
  holes}}, \href{http://dx.doi.org/10.1088/1126-6708/2004/04/048}{\emph{JHEP}
  {\bfseries 04} (2004) 048},
  [\href{https://arxiv.org/abs/hep-th/0401129}{{\ttfamily hep-th/0401129}}].

\end{thebibliography}\endgroup
\end{document}